\documentclass[letter,10pt]{article}
\usepackage{pstricks, pst-plot, pst-3d, pst-3dplot, multido, xkeyval, pst-light3d}

\usepackage[hang,small]{caption}
\usepackage[square,numbers,sort&compress]{natbib}
\usepackage{amsmath} 
\usepackage{amsfonts}
\usepackage{mathrsfs}

\usepackage{enumerate}
\usepackage{appendix}

\usepackage{epstopdf}
\usepackage[xdvi]{graphicx}
\usepackage{float} 
\graphicspath{pic/}

\usepackage{multicol}

\begin{document}
\newcommand{\sub}[2]{\mbox{$#1_\textrm{\scriptsize \hspace{0pt}#2}$}}
\newcommand{\subx}[2]{\mbox{$#1_\textrm{\tiny \hspace{0pt}#2}$}}
\newcommand{\suby}[2]{\mbox{${\scriptstyle #1}_\textrm{\scriptsize \hspace{0pt}#2}$}}
\newcommand{\subxy}[2]{\mbox{${\scriptstyle #1}_\textrm{\tiny \hspace{0pt}#2}$}}
\newcommand{\suballmath}[2]{\mbox{$#1_{\scriptstyle #2}$}}
\newcommand{\subxallmath}[2]{\mbox{$#1_{\scriptscriptstyle #2}$}}

\newcommand{\super}[2]{\mbox{$#1^\textrm{\scriptsize \hspace{0pt}#2}$}}
\newcommand{\superx}[2]{\mbox{$#1^\textrm{\tiny \hspace{0pt}#2}$}}

\newcommand{\subsuper}[3]{\mbox{$#1_\textrm{\scriptsize \hspace{0pt}#2}^\textrm{\scriptsize \hspace{0pt}#3}$}}
\newcommand{\subsuperx}[3]{\mbox{${\Large#1}_\textrm{\tiny \hspace{0pt}#2}^\textrm{\tiny \hspace{0pt}#3}$}}

\newcommand{\submathsuper}[3]{\mbox{$#1_{\scriptstyle #2}^\textrm{\scriptsize \hspace{0pt}#3}$}}
\newcommand{\submathsuperx}[3]{\mbox{$#1_{\scriptstyle #2}^\textrm{\tiny \hspace{0pt}#3}$}}
\newcommand{\subsuperallmath}[3]{\mbox{$#1_{\scriptstyle #2}^{\scriptstyle #3}$}}
\newcommand{\subsuperxallmath}[3]{\mbox{$#1_{\scriptscriptstyle #2}^{\scriptscriptstyle #3}$}}

\newcommand{\ziepsioverkt}[1]{\mbox{\normalsize $\frac{\sub{z}{#1}e\psi(\small \text{\emph{R}}_{\scriptscriptstyle{iM}})}{\normalsize {k}_{\tiny \text{B}} T}$}}
\newcommand{\angstrom}{\mbox{\text{\AA}}}

\newenvironment{new_enum}{
\begin{enumerate}
  \setlength{\itemsep}{1pt}
  \setlength{\parskip}{0pt}
  \setlength{\parsep}{0pt}
}{\end{enumerate}}

\newenvironment{new_item}{
\begin{itemize}
  \setlength{\itemsep}{1pt}
  \setlength{\parskip}{1pt}
  \setlength{\parsep}{-1pt}
}{\end{itemize}}

\title{Discrete Charge Effects on an Infinitely Long Cylindrical Rod Model}
\author{Ahmad A. J Agung{\small $^a$}, Christopher G. Jesudason {\small $^{a,b}$} \\
\small $^{a}$Department of Chemistry, University of Malaya, 50603 Kuala Lumpur, Malaysia\\
\small $^{b}$corresponding author: jesu@um.edu.my}
\date{}

\maketitle
\begin{abstract}
Two methods for determining the potential ($\psi$) around a discretely charged rod have been devised. The methods utilize the potential around the continuously charged rod $(\bar{\psi})$ as the reference where $\bar{\psi}$ is determined by the Poisson-Boltzmann equation. The potential data are used to determine the theoretical radial distribution function (RDF) which is compared with MD simulation data. It is shown that the magnitude of the charge and size parameters very strongly affects the shape of the RDF's and consequently the thermodynamics.
\end{abstract}

\setlength{\parindent}{2pt}
\section{Introduction}
This work is motivated by the oscillations that occur at the polyelectrolyte (PE) radial distribution function (RDF) of a system simulation (see Fig. \ref{fig:rdf-DNA}). The PE chain in that figure are our model for the DNA polyelectrolyte molecule. We use two kinds of PE models for the anionic bonded segment consisting of monomers of charge $\subx{z}{P}$. The first model has finite monomer charge $\subx{z}{P}=-12$ where the distance between adjacent monomers is $b=20.4~\angstrom$ and the second model has $\subx{z}{P}=-1$ with $b=1.7~\angstrom$. These models have the same line charge density $\delta_l=\subx{z}{P}/b$. These models have the same radius and charge density as the B-DNA chain \citep{Sin94}. The hard core radii of the DNA monomer is 8 $\angstrom$ and the soft core radii is 2 $\angstrom$. For the temperatures set in the simulation, the effective DNA radius can be taken as 10 $\angstrom$ because of the strength of the soft core repulsive potential which is effective over a distance of 2 $\angstrom$. Clearly we can observe oscillations at the RDF of the first model which does not appear in the second model. Thus the choice of the magnitude of the finite charge assigned to a model clearly affects the surrounding particle distribution. The degree of change of the RDF profiles with changing charge magnitude is very significant. In order to monitor these significant changes, here we focus on the simple model of an infinitely long charged rod where initially we focus charges being discrete and uniformly distributed along the rod. From the solution of the Poisson-Boltzmann equation (PBE) for the continuous (non-discrete) charge distribution of a charged rod, we can in at least two ways from the potential of the uniformly charged rod determine the potential field for a rod consisting of discrete charges. For a given (line) charge density $\delta_l$, the charge magnitude and intermolecular distance are arbitrary. In order to achieve the above results, we first provide a theoretical analysis and continue with the numerical solution of the nonlinear Poisson-Boltzmann equation (PBE) for this rod model. To substantiate our theoretical analysis, we perform molecular dynamics (MD) simulation of this rod model of discrete charges which is surrounded by a salt solution.\\
From the potential of a discretely charged rod, we determine theoretically the particle RDF about the rod. The method utilizes the Boltzmann factor $\sub{g}{f}(r)= \exp(\frac{-z_se\psi(r)}{\sub{k}{B}T})$ where $\frac{\rho_s}{\rho_{s,\infty}} = \sub{g}{f}$ is the ratio of the number density of particle $s$ at a distance $r$ over the bulk number density; $\psi(r)$ is the potential of average force \citep{Pry66}. This study is a first step in developing a model of calculating the general potential due to a discrete charge distribution enclosed by a boundary of arbitrary geometry. A principal motivation in attempting to provide a rigorous theory is due to the general tendency of workers to assume that the dimensions and parameters chosen from physical considerations for simulations is a fair representation of reality \cite{Hri00}; Our work shows that these choices are rather arbitrary and possibly inaccurate, which warrants a separate study of the connection between molecular size and the general potentials, since any one average potential of force $\psi(r)$ uniquely determines the RDF profiles.\\
\section{Theoretical Approach and Numerical Solution of PBE}
In the following we proceed to determine the non-uniform electric field distribution on a cylindrical Gaussian surface surrounding a charged rod. Then we relate it to the potential $\psi(r)$ and the particle distribution $g(r)$.\\
\subsection{Nonuniform Electric Field Around a Charged Rod}
\begin{figure}[h!]
\begin{pspicture}(-4.0,-0.2)(3,8.5)
\psset{xunit=0.6cm, yunit=0.6cm}

\psline[linecolor=black!70,linewidth=0.9pt, linestyle=dotted]{->}(-5.5,4.0)(20,4.0)
\psline[linecolor=black!80,linewidth=2pt]{-}(-0.6,4.0)(15.6,4.0)
\psline[linecolor=black!80,linewidth=2pt]{-}(-1.2,4.0)(-4.5,4.0)
\psline[linecolor=black!40,linewidth=0.7pt]{-}(-1.2,4.0)(-1.125,3.7)(-0.975,4.3)(-0.825,3.7)(-0.675,4.3)(-0.6,4.0)
\psline[linecolor=black!80,linewidth=2pt]{-}(16.2,4.0)(19.5,4.0)
\psline[linecolor=black!40,linewidth=0.7pt]{-}(15.6,4.0)(15.675,4.3)(15.825,3.7)(15.975,4.3)(16.125,3.7)(16.2,4.0)
\psdot[dotscale=1.8](0,4)
\psdot[dotscale=1.8](5,4)
\psdot[dotscale=1.8](10,4)
\psdot[dotscale=1.8](15,4)
\psdot[dotscale=1.8](-4.0,4)
\psdot[dotscale=1.8](19.0,4)

\psline[linecolor=black!70,linewidth=0.9pt, linestyle=dotted]{->}(-4.9,2.0)(-4.9,12.3)
\psline[linecolor=black!90,linewidth=0.5pt, linestyle=dashed](0,1.5)(0,4.3)
\psline[linecolor=black!90,linewidth=0.7pt](5,1)(5,9.5)
\psline[linecolor=black!90,linewidth=0.7pt](10,1)(10,9.5)
\psline[linecolor=black!90,linewidth=0.5pt, linestyle=dashed](15,1.5)(15,4.3)
\psline[linecolor=black!90,linewidth=0.5pt, linestyle=dashed](7,4)(7,8)
\psline[linecolor=black!90,linewidth=0.15pt]{|-|}(4.5,4.2)(4.5,8.0)
\rput(4.1,5.5){$r$}

\psline[linecolor=black!90,linewidth=0.7pt, linestyle=dashed]{-}(5,8)(10,8)
\psline[linecolor=black!90,linewidth=0.2pt]{|-|}(0.2,2.0)(4.8,2.0)
\psline[linecolor=black!90,linewidth=0.2pt]{|-|}(5.2,2.0)(9.8,2.0)
\psline[linecolor=black!90,linewidth=0.2pt]{|-|}(10.2,2.0)(14.8,2.0)
\rput(2.5,2.45){\small $b$}
\rput(7.5,2.45){\small $b$}
\rput(12.5,2.45){\small $b$}
\rput(19.5,5.0){${\mathrm{x}}$}
\rput(-4.0,13.0){${\mathrm{y}}$}
\pscurve[linecolor=black!90,linewidth=0.8pt]{->}(-3.8,5.0)(-4.0,5.2)(-4.5,4.7)(-4.6,4.0)(-4.4,3.1)(-4.3,2.9)(-4.0,2.7)(-3.9,3.1)
\rput(-4.0,2.2){${\mathrm{\theta}}$}

\psline[linecolor=black!70,linewidth=0.9pt, linestyle=dotted]{->}(7,8)(13,8)
\psline[linecolor=black!70,linewidth=0.9pt, linestyle=dotted]{->}(7,8)(7,14)
\rput(12.5,8.5){\small ${\mathrm{x}}^\prime$}
\rput(6.5,13.5){\small ${\mathrm{y}}^\prime$}
\psline[linecolor=black!90,linewidth=0.7pt]{->}(7,8)(9.2,10.6)
\psline[linecolor=black!90,linewidth=0.7pt, linestyle=dashed](5,4)(7,8)
\psline[linecolor=black!90,linewidth=1.0pt, linestyle=dotted](9.2,10.6)(7.5,12.1)
\psline[linecolor=black!90,linewidth=0.7pt]{->}(7,8)(5.3,9.4)
\psline[linecolor=black!90,linewidth=0.7pt, linestyle=dashed](10,4)(7,8)
\psline[linecolor=black!90,linewidth=1.0pt, linestyle=dotted](5.3,9.4)(7.5,12.1)
\psline[linecolor=black!90,linewidth=1.0pt]{->}(7,8)(7.5,12.1)
\psline[linecolor=black!70,linewidth=0.15pt, linestyle=dashed](-4.0,4)(7,8)
\psline[linecolor=black!70,linewidth=0.15pt, linestyle=dashed](19.0,4)(7,8)
\rput(0.4,6.0){\small $r_{1,j}$}
\rput(14.4,5.9){\small $r_{2,j}$}
\rput(-2.15,4.4){$\alpha_j$}
\psarc[linecolor=black!90,linewidth=0.4pt](-4.0,4){1.5}{0}{20}
\rput(17.2,4.4){$\beta_j$}
\psarc[linecolor=black!90,linewidth=0.4pt](19.0,4){1.5}{160}{180}
\pscircle(7,8){0.2}
\rput(9.5,10.9){\small $\mathbf{E}_1$}
\rput(5.0,9.9){\small $\mathbf{E}_2$}
\rput(7.4,12.5){$\mathbf{E}$}
\rput(7.5,8.17){$\mathbf{c}$}
\psline[linecolor=black!70,linewidth=0.1pt]{|-}(5.05,3.75)(5.7,3.75)
\rput(6,3.6){$p$}
\psline[linecolor=black!70,linewidth=0.1pt]{-|}(6.3,3.75)(6.95,3.75)
\psline[linecolor=black!70,linewidth=0.1pt]{|-}(7.05,3.75)(8.3,3.75)
\rput(8.6,3.6){$s$}
\psline[linecolor=black!70,linewidth=0.1pt]{-|}(9.0,3.75)(9.95,3.75)
\rput(8.4,6.6){\small $r_{2,1}$}
\rput(5.7,6.2){\small $r_{1,1}$}
\rput(9.3,4.4){$\beta_1$}
\psarc[linecolor=black!90,linewidth=0.6pt](10,4){0.7}{130}{180}
\rput(5.8,4.4){$\alpha_1$}
\psarc[linecolor=black!90,linewidth=0.6pt](5,4){0.7}{0}{61}
\end{pspicture}
\caption{\label{fig:mon-periodic}The electric fields experienced at the point $c$ (open circle) with a distance $r$ normal from the polyelectrolyte rod axis. The closed circles represents the discrete charge distribution of the rod.}
\end{figure}
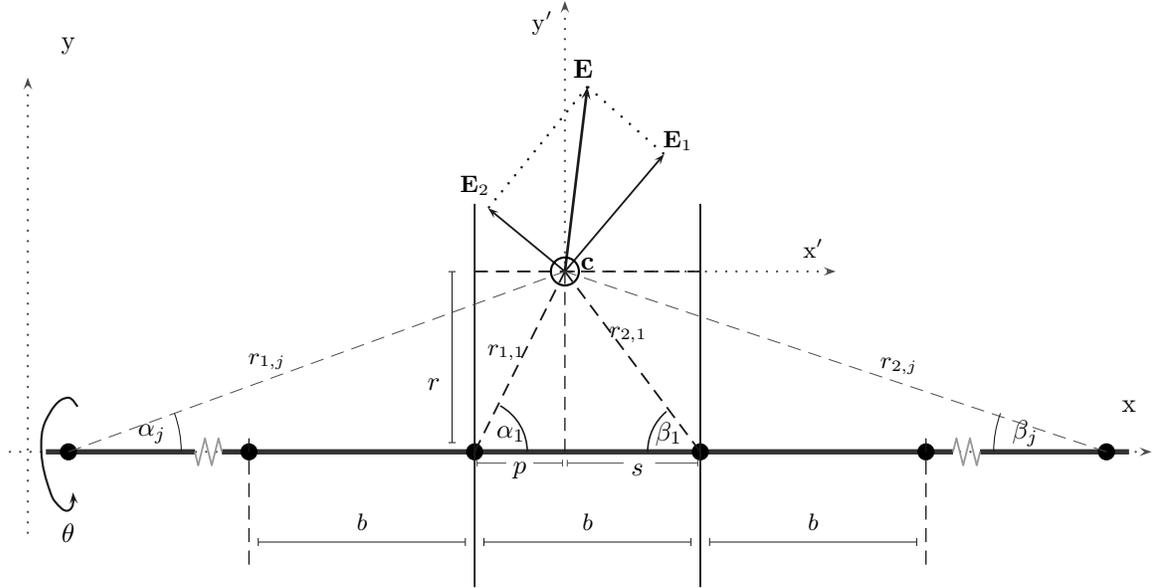

Fig. \ref{fig:mon-periodic} shows an array of charges (closed circles) along the rod axis. The rod is extended to infinity to the left and right. A test charge $c$ (open circle) lies at a perpendicular distance $r$ to the rod axis. We start by determining the net electric field at point $c$ due to the rod charges. For the moment, we assume no interference of counterion charges. We use bold letters to denote vectors and italic letters to represent its magnitude. In general, the first subscript in our variables refers to the left (value 1) half of the diagram and the subscript with value 2 refers to the right half of the same. The second subscript denotes the charge index along the rod in the appropriate direction indicated by the first subscript.\\
The angle $\alpha_j$ is the angle between the rod axis and a straight line from the point $c$ to the left part of the rod charges. The index $j$ starts from unity for the closest rod charge at the centre of the diagram and ends at infinity. $\mathbf{E}_{1,j}$ is the electric field at point $c$ contributed by the $j^{th}$ charge on the left of the rod. Analogous to $\alpha_j$ and $\mathbf{E}_{1,j}$, the angle $\beta_j$ and electric field $\mathbf{E}_{2,j}$ pertains to the angle and electric field for right part of the rod. In Fig. \ref{fig:mon-periodic}, if $r_{1,1}$ denotes the distance between the test charge $c$ and the nearest rod charge on the left side, $p$ is the projection of the $r_{1,1}$ line to the rod $x$ axis.  Similarly, $s$ is the projection of the line $r_{2,1}$ to the rod axis on the right hand side and $s=(b-p)$. For later use, we define
\begin{align}
p_j & = p + (j-1)b \\
s_j & = (b-p) + (j-1)b
\end{align}
Below is the net electric field (in cgs unit) $\mathbf{E}_1$ due to the charges on the left. 
\begin{align} 
\mathbf{E}_1 &= \sum_{j=1}^\infty \mathbf{E}_{1,j} = \sum_{j=1}^\infty \frac{\subx{z}{P}}{r_{1,j}^2}\,\mathbf{\hat{r^\prime}}_{1,j}  \label{eq:one} \text{ , }
\intertext{where $\mathbf{\hat{r^\prime}}_{1,j}$ is the unit vector from the $j^{th}$ charge to the test charge $c$ and $\sub{z}{p}$ is the magnitude of the discrete charges, or the total charge of a continuously charged rod over distance $b$. The prime coordinate are centred on test charge $c$, whereas the unprimed coordinates refer to the rod axis, where the zero for $x$ is arbitrary, and the zero for the $y$ coordinate is located on the rod. Components of $r^\prime$, $x^{\prime}$  and $y^{\prime}$, refer to orthogonal bases $\mathbf{\hat{x^\prime}}$ and $\mathbf{\hat{y^\prime}}$ that are parallel and perpendicular to the rod respectively. In mks units, we multiply by the coefficient $1/4\pi\epsilon$ before the summation of the (\ref{eq:one}). We use cgs unit here for reasons of convenience. We can separate each $\mathbf{E}_{1,j}$ to their horizontal ($x^\prime$) and vertical ($y^\prime$) components. Then $\mathbf{E}_1$ in terms of its $x^\prime$ and $y^\prime$ components are }
\mathbf{E}_{1,x^\prime} &= \sum_{j=1}^\infty \frac{\subx{z}{P}}{r_{1,j}^2}\, \cos\alpha_j \,\mathbf{\hat{x^\prime}} = \sum_{j=1}^\infty \frac{\subx{z}{P}}{r_{1,j}^3}\, p_j \,\mathbf{\hat{x^\prime}} \\
\mathbf{E}_{1,y^\prime} &= \sum_{j=1}^\infty \frac{\subx{z}{P}}{r_{1,j}^2}\, \sin \alpha_j \,\mathbf{\hat{y^\prime}} = \sum_{j=1}^\infty \frac{\subx{z}{P}}{r_{1,j}^3}\, r \,\mathbf{\hat{y^\prime}}
\end{align}
Similarly for the electric field $\mathbf{E}_2$ due to the charges at the RHS of the rod
\begin{align}
\mathbf{E}_2 &= \sum_{j=1}^\infty \mathbf{E}_{2,j} = \sum_{j=1}^\infty \frac{\subx{z}{P}}{r_{2,j}^2}\,\mathbf{\hat{r^\prime}}_{2,j}\\
\mathbf{E}_{2,x^\prime} &= - \sum_{j=1}^\infty \frac{\subx{z}{P}}{r_{2,j}^2} \cos\beta_j \,\mathbf{\hat{x^\prime}} = - \sum_{j=1}^\infty \frac{\subx{z}{P}}{r_{2,j}^3} s_j \,\mathbf{\hat{x^\prime}} \\
\mathbf{E}_{2,y^\prime} &= \sum_{j=1}^\infty \frac{\subx{z}{P}}{r_{2,j}^2} \sin \beta_j  \,\mathbf{\hat{y^\prime}} = \sum_{j=1}^\infty \frac{\subx{z}{P}}{r_{2,j}^3}\, r  \,\mathbf{\hat{y^\prime}} \text{ , }
\end{align}
where the negative sign in the $\mathbf{E}_{2,x}$ expression is due to the reverse polarity of the $x^\prime$ axis. Since $\mathbf{E}_1$ and $\mathbf{E}_2$ act at the same point $c$, we can directly sum the net electric field for the $x^\prime$ and $y^\prime$ directions, which are 
\begin{align} 
\mathbf{E}_{x^\prime} &= \mathbf{E}_{1,x^\prime} + \mathbf{E}_{2,x^\prime} =  \sum_{j=1}^\infty \subx{z}{P} \left ( \frac{p_j}{r_{1,j}^3} - \frac{s_j}{r_{2,j}^3} \right ) \,\mathbf{\hat{x^\prime}} \label{eq:Ex}\\
\mathbf{E}_{y^\prime} &= \mathbf{E}_{1,y^\prime} + \mathbf{E}_{2,y^\prime} = \sum_{j=1}^\infty \subx{z}{P} \left ( \frac{r}{r_{1,j}^3} + \frac{r}{r_{2,j}^3} \right ) \,\mathbf{\hat{y^\prime}} \text{  ,} \label{eq:Ey}
\end{align}
Then the magnitude of the resultant electric field acting at point $c$ becomes
\begin{align} 
\it{E} &= \sqrt{\mathit{E}_{x^\prime}^2 + \emph{E}_{y^\prime}^2} = \subx{z}{P} \left [ \left ( \sum_{j=1}^\infty \left ( \frac{p_j}{r_{1,j}^3} - \frac{s_j}{r_{2,j}^3} \right ) \right )^2 +  \left ( \sum_{j=1}^\infty \left ( \frac{r}{r_{1,j}^3} - \frac{r}{r_{2,j}^3} \right ) \right )^2 \right ]^{1/2}. \label{eq:res_E}
\end{align}
The $r_{1,j}$ and $r_{2,j}$ expressions are (see Fig. \ref{fig:mon-periodic})
\begin{align}
r_{1,j} &= \left[ {(p+(j-1)b)^2+r^2} \right ]^{1/2}\\
r_{2,j} &= \left [ {((b-p)+(j-1)b)^2+r^2} \right ]^{1/2} \text{~~.}
\end{align}
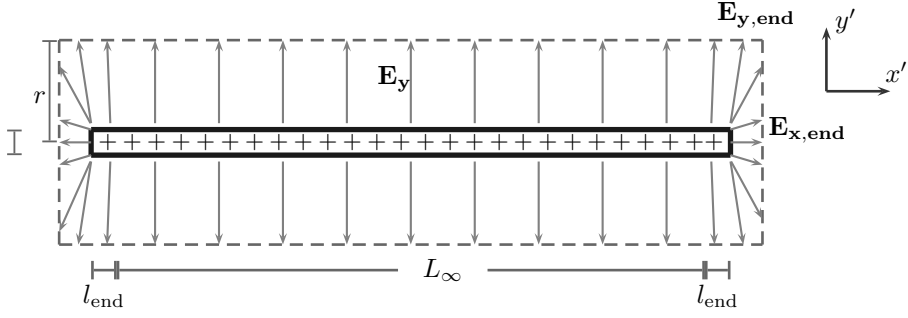
\begin{figure}[h!]
\begin{pspicture}(-4.2,-2.3)(5,2.2)
\psset{xunit = 0.85cm, yunit=0.85cm}

\psline[linewidth=2.0pt,linecolor=black!90]{-}(-3.0,0.4)(7.0,0.4)
\psline[linewidth=2.0pt,linecolor=black!90]{-}(-3.0,0.0)(7.0,0.0)
\psline[linewidth=2.0pt,linecolor=black!90]{-}(-3.0,0.4)(-3.0,0.0)
\psline[linewidth=2.0pt,linecolor=black!90]{-}(7.0,0.4)(7.0,0.0)
\rput(2.0,0.2){\small{$++++++++++++++++++++++++++$}}

\psline[linewidth=1.0pt,linecolor=black!60, linestyle=dashed]{-}(-3.5,1.8)(7.5,1.8)
\psline[linewidth=1.0pt,linecolor=black!60, linestyle=dashed]{-}(-3.5,-1.4)(7.5,-1.4)
\psline[linewidth=1.0pt,linecolor=black!60, linestyle=dashed]{-}(-3.5,-1.4)(-3.5,1.8)
\psline[linewidth=1.0pt,linecolor=black!60, linestyle=dashed]{-}(7.5,-1.4)(7.5,1.8)

\psline[linewidth=1.0pt,linecolor=black!60]{|-}(-2.6,-1.8)(1.8,-1.8)
\psline[linewidth=1.0pt,linecolor=black!60]{-|}(3.2,-1.8)(6.6,-1.8)
\rput(2.5,-1.8){$L_\infty$}

\psline[linewidth=1.0pt,linecolor=black!60]{|-|}(6.6,-1.8)(7.0,-1.8)
\rput(6.8,-2.2){$\sub{l}{end}$}

\psline[linewidth=1.0pt,linecolor=black!60]{|-|}(-2.6,-1.8)(-3.0,-1.8)
\rput(-2.8,-2.2){$\sub{l}{end}$}

\rput(8.2,0.4){$\mathbf{E_{x,end}}$}
\rput(7.4,2.2){$\mathbf{E_{y,end}}$}
\rput(1.75,1.2){$\mathbf{E_y}$}

\psline[linewidth=0.7pt,linecolor=black!60]{|-|}(-4.2,0.4)(-4.2,0)
\psline[linewidth=0.7pt,linecolor=black!60]{|-|}(-3.65,0.2)(-3.65,1.8)
\rput(-3.8, 0.9){$r$}

\psline[linewidth=1.0pt, linecolor=black!80]{->}(8.5,1.0)(9.5,1.0)
\psline[linewidth=1.0pt, linecolor=black!80]{->}(8.5,1.0)(8.5,2.0)
\rput(9.6, 1.3){$x^\prime$}
\rput(8.8, 2.1){$y^\prime$}

\psline[linewidth=0.8pt, linecolor=black!50]{->}(7.01,0.5)(7.5,1.4)
\psline[linewidth=0.8pt, linecolor=black!50]{->}(7.0,0.5)(7.2,1.8)
\psline[linewidth=0.8pt, linecolor=black!50]{->}(6.7,0.5)(6.75,1.8)
\psline[linewidth=0.8pt, linecolor=black!50]{->}(6.0,0.5)(6.0,1.8)
\psline[linewidth=0.8pt, linecolor=black!50]{->}(5.0,0.5)(5.0,1.8)
\psline[linewidth=0.8pt, linecolor=black!50]{->}(4.0,0.5)(4.0,1.8)
\psline[linewidth=0.8pt, linecolor=black!50]{->}(3.0,0.5)(3.0,1.8)
\psline[linewidth=0.8pt, linecolor=black!50]{->}(2.0,0.5)(2.0,1.8)
\psline[linewidth=0.8pt, linecolor=black!50]{->}(1.0,0.5)(1.0,1.8)
\psline[linewidth=0.8pt, linecolor=black!50]{->}(0.0,0.5)(0.0,1.8)
\psline[linewidth=0.8pt, linecolor=black!50]{->}(-1.0,0.5)(-1.0,1.8)
\psline[linewidth=0.8pt, linecolor=black!50]{->}(-2.0,0.5)(-2.0,1.8)
\psline[linewidth=0.8pt, linecolor=black!50]{->}(-2.7,0.5)(-2.75,1.8)
\psline[linewidth=0.8pt, linecolor=black!50]{->}(-3.0,0.5)(-3.2,1.8)
\psline[linewidth=0.8pt, linecolor=black!50]{->}(-3.01,0.5)(-3.5,1.4)

\psline[linewidth=0.8pt, linecolor=black!50]{->}(7.01,-0.1)(7.5,-1.0)
\psline[linewidth=0.8pt, linecolor=black!50]{->}(7.0,-0.1)(7.2,-1.4)
\psline[linewidth=0.8pt, linecolor=black!50]{->}(6.7,-0.1)(6.75,-1.4)
\psline[linewidth=0.8pt, linecolor=black!50]{->}(6.0,-0.1)(6.0,-1.4)
\psline[linewidth=0.8pt, linecolor=black!50]{->}(5.0,-0.1)(5.0,-1.4)
\psline[linewidth=0.8pt, linecolor=black!50]{->}(4.0,-0.1)(4.0,-1.4)
\psline[linewidth=0.8pt, linecolor=black!50]{->}(3.0,-0.1)(3.0,-1.4)
\psline[linewidth=0.8pt, linecolor=black!50]{->}(2.0,-0.1)(2.0,-1.4)
\psline[linewidth=0.8pt, linecolor=black!50]{->}(1.0,-0.1)(1.0,-1.4)
\psline[linewidth=0.8pt, linecolor=black!50]{->}(0.0,-0.1)(0.0,-1.4)
\psline[linewidth=0.8pt, linecolor=black!50]{->}(-1.0,-0.1)(-1.0,-1.4)
\psline[linewidth=0.8pt, linecolor=black!50]{->}(-2.0,-0.1)(-2.0,-1.4)
\psline[linewidth=0.8pt, linecolor=black!50]{->}(-2.7,-0.1)(-2.75,-1.4)
\psline[linewidth=0.8pt, linecolor=black!50]{->}(-3.0,-0.1)(-3.2,-1.4)
\psline[linewidth=0.8pt, linecolor=black!50]{->}(-3.01,-0.1)(-3.5,-1.2)

\psline[linewidth=0.8pt, linecolor=black!50]{->}(7.0,0.4)(7.5,0.55)
\psline[linewidth=0.8pt, linecolor=black!50]{->}(7.0,0.2)(7.5,0.2)
\psline[linewidth=0.8pt, linecolor=black!50]{->}(7.0,0.0)(7.5,-0.15)

\psline[linewidth=0.8pt, linecolor=black!50]{->}(-3.0,0.4)(-3.5,0.55)
\psline[linewidth=0.8pt, linecolor=black!50]{->}(-3.0,0.2)(-3.5,0.2)
\psline[linewidth=0.8pt, linecolor=black!50]{->}(-3.0,0.0)(-3.5,-0.15)

\end{pspicture}
\caption{\label{fig:neglect-ends-cyl}Infinitely charged long rod (solid lines) inside a cylindrical closed surface (dashed lines). The continuous line charge density of the rod is $\delta_l$.}
\end{figure}
Fig. \ref{fig:neglect-ends-cyl} depicts a continuously charged rod inside a closed cylindrical Gaussian surface. The figure uses ($x^\prime, y^\prime$) coordinate. Except for the electric field near the two ends of the rod, the direction of electric field is always perpendicular to the rod axis since the components parallel to the rod axis cancel everywhere. We define $\sub{l}{end}$ as the length at the two ends of the rod where the electric field direction is not perpendicular to the rod axis. The $\sub{l}{end}$ length is approximately finite depending on the rod charge and dimension. We define $L_\infty$ as the length within the rod axis where the electric field directions are perpendicular to the rod axis. Since the rod has infinite length and $\sub{l}{end}$ is finite, obviously $L_\infty$  limits at infinity. From Gauss' law, the relationship between the electric field at the Gaussian surface and the total rod charge $\sub{Q}{Rod}$ is \cite{Sch87}
\begin{align}
\int_S \mathbf{E}\cdot d\mathbf{A} & = 4\pi\,\subx{Q}{Rod} \label{eq:lor} 
\end{align}
Breaking (\ref{eq:lor}) into the surface components of the closed cylindrical surface in Fig. \ref{fig:neglect-ends-cyl}, we obtain
\begin{equation}
E_y\, 2\pi r L_\infty + 2 \int_0^{l_{end}} \sub{E}{y,end}(x^\prime,y^\prime)\,2\pi r\, \mathrm{d}l + 2 \int_0^r \sub{E}{x,end}(x^\prime,y^\prime)\, 2\pi r_i\, \mathrm{d}r_i = 4 \pi L_\infty\, \delta_l \label{eq:wlw}
\end{equation}
where $E_y$ is the electric field magnitude within the range $L_\infty$ where the direction is perpendicular to the rod axis. $\sub{E}{y,end}$ and $\sub{E}{x,end}$ are the electric field components from the $\sub{l}{end}$ areas where the directions are parallel to the $y^\prime$ and $x^\prime$ axes respectively. $\delta_l$ is the rod line charge density. Divide all terms in (\ref{eq:wlw}) by $2\pi\,r\,L_\infty$, we obtain
\begin{equation}
E_y + \frac{2 \int_0^{l_{end}} \sub{E}{y,end}(x^\prime,y^\prime)\, \mathrm{d}l}{L_\infty} + \frac{2 \int_0^r \sub{E}{x,end}(x^\prime,y^\prime)\, r_i\, \mathrm{d}r_i}{r\,L_\infty} = \frac{2\, \delta_l}{r} \label{eq:wlu}
\end{equation}
Since the $\sub{l}{end}$ is finite, the numerators at the second and the third terms in the LHS of (\ref{eq:wlu}) are also finite. Thus as $L_\infty \longrightarrow \infty$, 
\begin{equation}
E_y \cong \frac{2\, \delta_l}{r} \label{eq:wlp} \text{ , }
\end{equation}
We define ${\Bar{E}}= E_y$ as the electric field due to a continuously charged rod. Henceforth any barred symbols denote properties of a continuously charged rod model.\\
If $z_p$ is the total finite charge within an axial distance $b$ of a continuously charged rod, then $z_p = b\,\delta_l$ and
\begin{equation}
\mathit{\bar{E}} = \frac{2\,\delta_l\,b}{r\,b} = \frac{2\subx{z}{P}}{r\,b} \text{ .} \label{eq:E-bar} 
\end{equation}
If we define 
\begin{align}
D_x(,b,r) &= \frac{rb}{2} \sum_{j=1}^\infty \left(  \frac{p_j}{r_{1,j}^3} - \frac{s_j}{r_{2,j}^3} \right ) \text{} \label{eq:Dx} \\
D_y(p,b,r) &= \frac{rb}{2} \sum_{j=1}^\infty \left ( \frac{r}{r_{1,j}^3} + \frac{r}{r_{2,j}^3}  \right ) \text{ } \label{eq:Dy}\\
D(p,b,r) &= \frac{rb}{2} \left [ \left( \sum_{j=1}^\infty \left(  \frac{p_j}{r_{1,j}^3} - \frac{s_j}{r_{2,j}^3} \right ) \right )^2 + \left ( \sum_{j=1}^\infty \left ( \frac{r}{r_{1,j}^3} + \frac{r}{r_{2,j}^3}  \right )  \right )^2 \right ]^{1/2}\text{ , } \label{eq:D}
\intertext{then by (\ref{eq:Ex}$\--$\ref{eq:res_E}) and (\ref{eq:E-bar}$\--$\ref{eq:D}), the following equalities hold} 
\mathit{E_x} &= D_x\,\mathit{\bar{E}} \label{eq:DxE} \text{  }\\
\mathit{E_y} &= D_y\,\mathit{\bar{E}} \label{eq:DyE} \text{  }\\
\mathit{E} &= D\,\mathit{\bar{E}} \label{eq:DE} \text{ . }
\end{align}
We define $D$ as the electrostatic field distribution function.\\
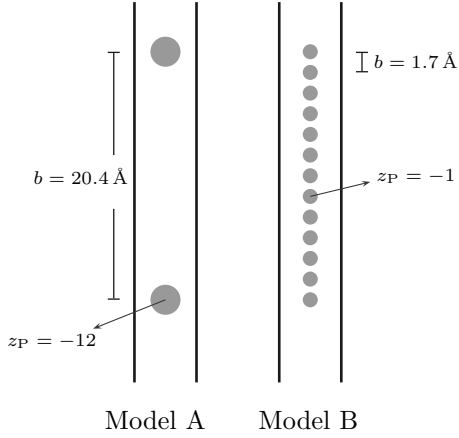
\begin{figure}[h!]
\begin{pspicture}(-1.8,-1.0)(10,5.0)
\psset{xunit=0.55cm, yunit=0.55cm}

\psline[linecolor=black!90,linewidth=1.0pt](1.5,-0.5)(1.5,8.7)
\psline[linecolor=black!90,linewidth=1.0pt](3.0,-0.5)(3.0,8.7)
\rput(2,-1.4){Model A}
\pscircle*[linecolor=black!40, linewidth=0.5pt](2.25,1.5){0.2}
\pscircle*[linecolor=black!40, linewidth=0.5pt](2.25,7.5){0.2}
\psline[linecolor=black!90,linewidth=0.5pt]{|-}(1.0,1.5)(1.0,3.7)
\psline[linecolor=black!90,linewidth=0.5pt]{-|}(1.0,5.0)(1.0,7.5)
\rput(0.2,4.5){\scriptsize{$b = 20.4\,\angstrom$}}
\psline[linecolor=black!80, linewidth=0.4pt]{->}(2.25,1.5)(0.5,0.8)
\rput(-0.5,0.5){\scriptsize{$\subx{z}{P} = -12$}}

\psline[linecolor=black!90,linewidth=1.0pt](5.0,-0.5)(5.0,8.7)
\psline[linecolor=black!90,linewidth=1.0pt](6.5,-0.5)(6.5,8.7)
\rput(5.7,-1.4){Model B}
\pscircle*[linecolor=black!40, linewidth=0.5pt](5.75,1.5){0.1}
\pscircle*[linecolor=black!40, linewidth=0.5pt](5.75,2.0){0.1}
\pscircle*[linecolor=black!40, linewidth=0.5pt](5.75,2.5){0.1}
\pscircle*[linecolor=black!40, linewidth=0.5pt](5.75,3.0){0.1}
\pscircle*[linecolor=black!40, linewidth=0.5pt](5.75,3.5){0.1}
\pscircle*[linecolor=black!40, linewidth=0.5pt](5.75,4.0){0.1}
\pscircle*[linecolor=black!40, linewidth=0.5pt](5.75,4.5){0.1}
\pscircle*[linecolor=black!40, linewidth=0.5pt](5.75,5.0){0.1}
\pscircle*[linecolor=black!40, linewidth=0.5pt](5.75,5.5){0.1}
\pscircle*[linecolor=black!40, linewidth=0.5pt](5.75,6.0){0.1}
\pscircle*[linecolor=black!40, linewidth=0.5pt](5.75,6.5){0.1}
\pscircle*[linecolor=black!40, linewidth=0.5pt](5.75,7.0){0.1}
\pscircle*[linecolor=black!40, linewidth=0.5pt](5.75,7.5){0.1}
\psline[linecolor=black!90,linewidth=0.5pt]{|-|}(7.0,7.0)(7.0,7.5)
\rput(8.3,7.3){\scriptsize{$b = 1.7\,\angstrom$}}
\rput(8.3,4.5){\scriptsize{$\subx{z}{P} = -1$}}
\psline[linecolor=black!80, linewidth=0.4pt]{->}(5.75,4.0)(7.2,4.35)

\end{pspicture}
\caption{\label{fig:rod-models}Two charged rod models studied. Both model has the same line charged density $\subx{z}{P}/b$}
\end{figure}
We utilize two kinds of discretely charged rod models shown in Fig. \ref{fig:rod-models} with the same line and volume charge density. Figure \ref{fig:normal-Dy} graphs the function $D_y=\mathit{E_y}/\mathit{\bar{E}}$ for the two models. Fig. \ref{fig:normal-Dy}.a for the rod model A whose finite charge $-12$ and Fig. \ref{fig:normal-Dy}.b for model B whose finite charge $-1$. $D_y$ is determined numerically by extending $j$ to a large number ($\sim 5,000$) until no change is observed. From the figures, the electrostatic field distribution around the rod model B is rather uniform. For the rod model A the degree of variation of the curves for different $r$ values is very obvious. The degree of variation diminishes as the perpendicular distance $r$ increases.\\
\begin{figure}[h!]
\begin{minipage}[b]{8cm}
\includegraphics[scale=0.7]{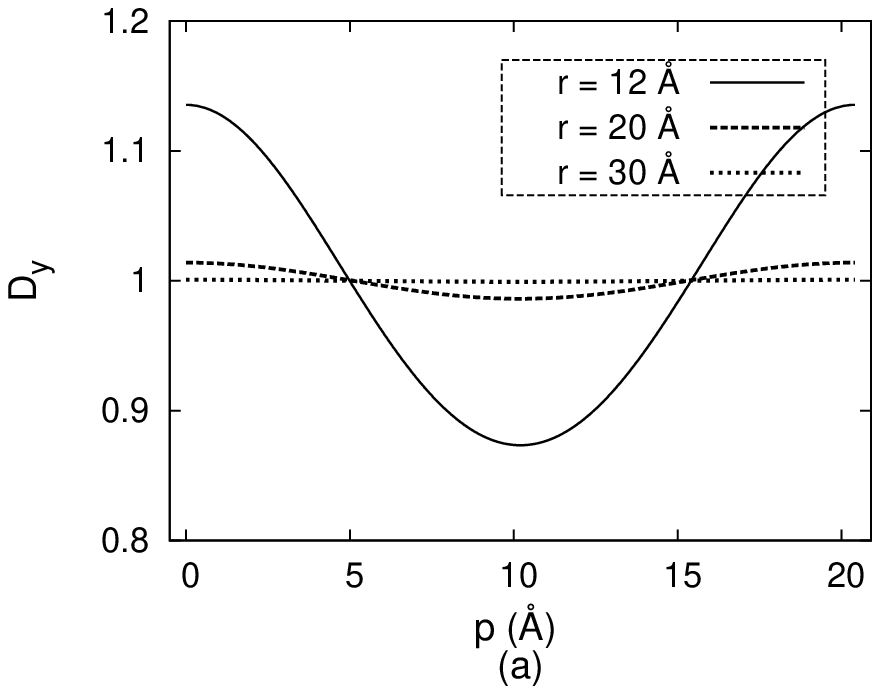}
\end{minipage}
\hspace{0.2cm}
\begin{minipage}[b]{8cm}
\includegraphics[scale=0.7]{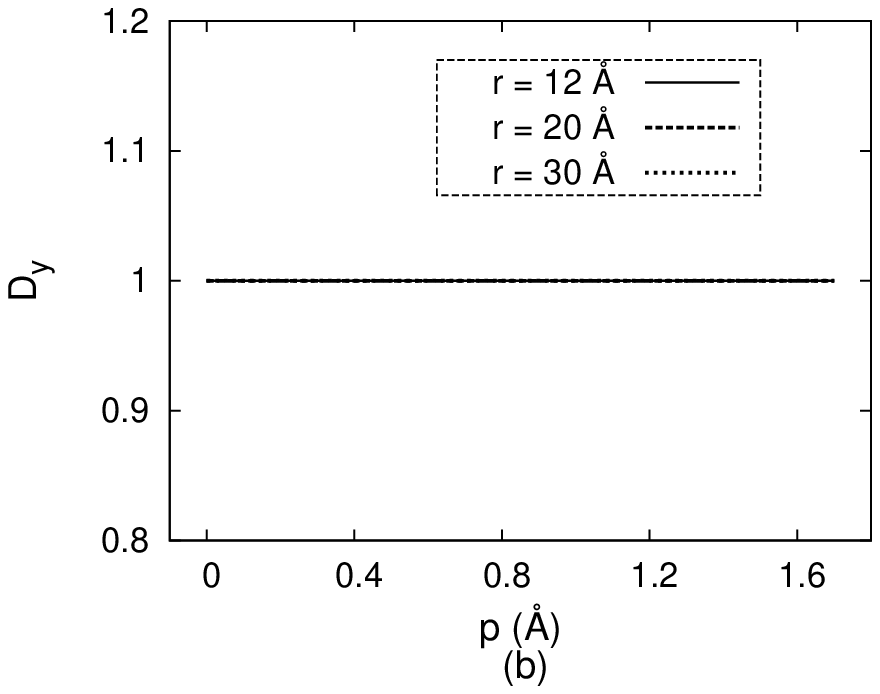}
\end{minipage}
\caption{\label{fig:normal-Dy} Function $D_y(p) = \mathit{E_y}/\mathit{\bar{E}}$ (\ref{eq:DyE}) for: (Fig. a) rod model A ($\subx{z}{P}=-12$ and $b=20.4 \angstrom$) and (Fig. b) rod model B ($\subx{z}{P}=-1$ and $b=1.7 \angstrom$). $r$ is the perpendicular distance to the rod axis.}
\end{figure}
Figure \ref{fig:normal-Dx} graphs the $|D_x(p)| = \left | \mathit{E_x}/\mathit{\bar{E}} \right |$ for model A and model B. Like $D_y$, the variation of $D_x$ appears only for model A and the variation vanishes as the perpendicular distance $r$ increases. From Fig. \ref{fig:normal-Dy} and \ref{fig:normal-Dx}, the values of $|D_x|$ in general are much less than $D_y$. Since $D=\sqrt{D_x^2+ D_y^2}$ (from (\ref{eq:Dx}-\ref{eq:D})) the contribution of $D_y$ to $D$ is very much larger compared to the $D_x$ contribution.
\begin{figure}[h!]
\begin{minipage}[b]{8cm}
\includegraphics[scale=0.7]{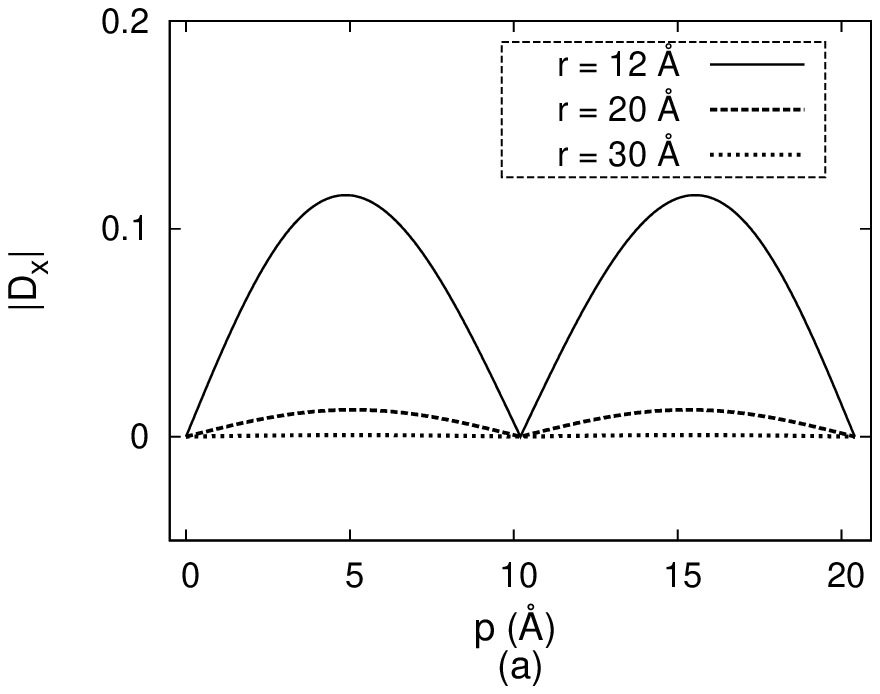}
\end{minipage}
\hspace{0.2cm}
\begin{minipage}[b]{8cm}
\includegraphics[scale=0.7]{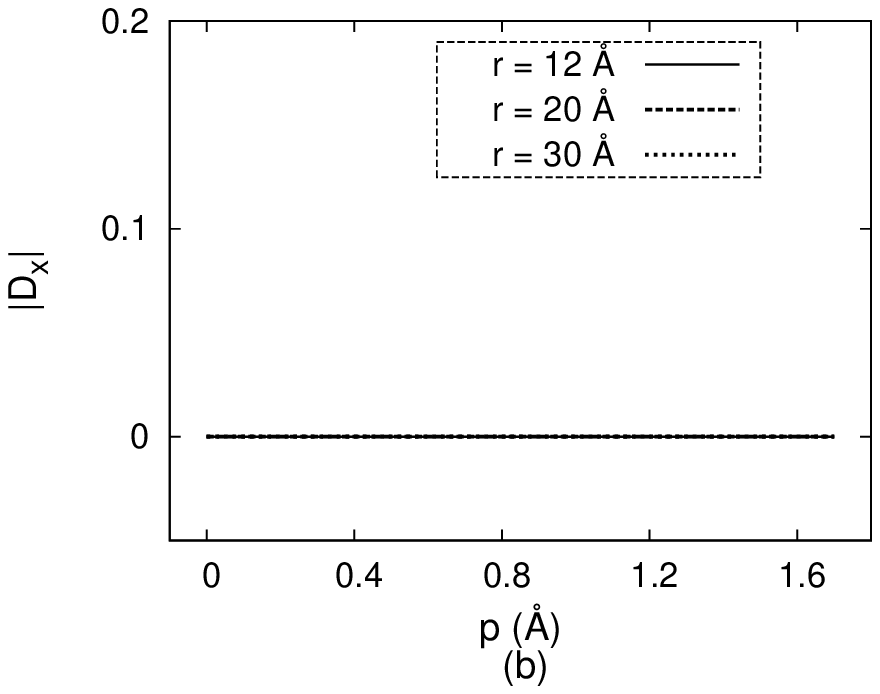}
\end{minipage}
\caption{\label{fig:normal-Dx} Function $|D_x(p)| = \left| \mathit{E_x}/\mathit{\bar{E}} \right | $ (\ref{eq:DxE}) for: (Fig. a) rod model A ($\subx{z}{P}=-12$ and $b=20.4 \angstrom$) and (Fig. b) rod model B ($\subx{z}{P}=-1$ and $b=1.7 \angstrom$). $r$ is the perpendicular distance to the rod axis.}
\end{figure}

In the following we propose a model when the counterion and salt ions are included in the system containing an infinitely charged rod. If the discrete rod charges along the rod axis have equal axial charge distance $b$, the ionic distribution about any two tangential plane that are equivalent by symmetry must be the same. We note that the $D$ function i not affected by the counterion about the rod. Hence we may couple $D$ to the effective dielectric constant that arise from the water and ionic distribution about the rod to relate $E$ and $\bar{E},$.\\
\subsection{Multigrid Method for Solving the PBE}\label{subsec:MultPBE}
For some simple and symmetric shapes (e.g. plate, cylinder or sphere), the PBE expressions are well-known in their second order differential equation forms. The linearized PBE known as the Debye-Huckel approximation provides an analytical expressions that holds only for low potentials. As far as the second order differential form of the PBE is concerned, most solutions need to be found numerically. In this work we solve the PBE for both the continuous and discrete charged rod model. The potential for the continuous model will be used as a means to obtain the potential of the discretely charged rod by methods described in the two next subsections.\\
For a continuous charged rod, the PBE is expressed as
\begin{equation}
\frac{\partial^2\bar{\phi}(r)}{\partial r^2} + \frac{1}{r}\left (\frac{\partial\bar{\phi}(r)}{\partial r}\right) - \sum_{s=1}^n A_s\exp(-z_s\bar{\phi}(r)) = 0, \label{eq:cylPB}
\end{equation}
where $\bar{\phi}(r)=e\bar{\psi}(r)/\sub{k}{B}T$ is the reduced potential and $A_s= -e^2z_s\rho_{s,\infty}/(\epsilon \sub{k}{B}T)$. The Neumann boundary conditions are obtained by applying Gauss' law at the rod-counterion contact distance $r=\sub{r}{p}=(\subx{a}{P}+\sub{r}{cion})$ and $r=\sub{r}{c}$ as follow
\begin{align}
\frac{\partial\bar{\phi}(r=\subx{a}{P}+\sub{r}{cion})}{\partial r} & =-\frac{\subx{z}{P}e^2}{\epsilon 2\pi(\subx{a}{P}+\sub{r}{cion})b\sub{k}{B}T}  \quad \text{and} \\
\quad \frac{\partial\bar{\phi}(r=\sub{r}{c})}{\partial r} & =0~ \text { ,} \label{eq:2nd-bound}
\end{align}
where $\subx{a}{P}$ is the rod radius and $\sub{r}{cion}$ is the counterion radius. \subx{z}{P} is the magnitude of the individual charge of the rod in the discrete case or the total charge contained within a distance $b$ (for continuous rod). The cylinder cell radius \sub{r}{c} is determined by a non-local macroscopic rod charge density $\sub{\rho}{\sub{Z}{rod}}$ (per unit volume). where\\
\begin{align}
\sub{\rho}{\sub{Z}{Rod}} & = \frac{\subx{Z}{Rod}}{\subx{V}{Simulation box}} = \frac{\subx{Z}{Rod}}{\subx{V}{Cylinder PBE box}} \text{ ~~,}
\end{align}
where $\sub{Z}{Rod}$ is the total charge of the rod in one simulation box; $\subx{V}{Simulation box}=\sub{L}{Box}^3$ is the cubic simulation box volume with simulation box length $\sub{L}{Box}$ which is set (see Section \ref{sec:sim}). Since the length of the charged rod (with total charge $\sub{Z}{Rod}$) in one simulation box equals $\sub{L}{Box}$, the length of the cylindrical PBE per $\sub{Z}{Rod}$ charge also equals $\sub{L}{Box}$. Thus
\begin{align}
\frac{1}{\sub{L}{Box}^3} & = \frac{1}{\sub{L}{Box} \pi r_c^2}  \text{~ or}\\
r_c & = \sqrt{\frac{1}{\pi}} \sub{L}{Box} 
\end{align}
The second boundary condition (Eq. (\ref{eq:2nd-bound})) is applied for at least three reasons. First, at large distances from the rod axis, the potential is about constant implying zero potential gradient. Second, due to charge neutrality of the cylindrical box, theoretically the surface charge density equals zero at the box surface leading to zero electric field. Third, the systems studied in the theoretical PBE are substantiated by simulations utilizing the periodic boundary condition (PBC). If the same model like PBC is applied in PBE calculation, there is a mirroring shape of the potential between two neighboring boxes (in the PBE calculation). We assume that our simulation box is large enough to warrant (\ref{eq:2nd-bound}).\\
The multigrid method applied in this research uses the Full Approximation Storage (FAS) procedure described in \cite{Bra77}. The outline of the algorithm for our application is given in Appendix \ref{appA} based on a slight modification of the equations used in the code given in \citep{Pre92} which was developed on the basis of reference \citep{Bra77}. The way we differ follows. Reference \citep{Pre92} uses another equation, and not (\ref{eq:cylPB}). Further, the code in \citep{Pre92} is in 2-D with uniform and equal number of grid for both axes, whereas ours is both in 1 and 2-D where in 2-D the number of grid for both axes can be different; \citep{Pre92} uses the Newton-Gauss-Seidel method for the smoothing procedures and we utilize the globally convergent Newton-Raphson and conjugate gradient method. Another difference is \citep{Pre92} uses the Dirichlet boundary conditions, whereas we use the Neumann boundary conditions. \citet{Obe93} and \citet{Hol95} have also used multigrid methods to obtain the solution of the PBE. We had examined the uniqueness of the PBE solution by supplying different initial conditions that led to the same final solution. The convergence criteria is taken when the residual Euclidean norm of (\ref{eq:cylPB}) is zero. Computationally, the iterations fluctuate about zero and we choose a tolerance number ($\sim 2\times 10^{-6}$) of this norm that define our numerical solution. The physical variables used for our theoretical PBE calculations are exactly the same as for our MD simulations, so that the results from MD may be compared to the theoretical model. The detail of the physical variables of the MD systems and the theoretical models are given in Section \ref{sec:sim}.
\subsection{Obtaining the Potential About a Discretely Charged Rod} 
After examining the electric field around the rod axis, we relate it to the potential. Often, the scalar potential is the starting point for analysis and we adopt this route. We propose two methods to determine the non-uniform potential around a discretely charged rod. Both methods require the potential data around a continuously charged rod ($\bar{\phi}$). The first method directly modifies the $\bar{\phi}$ by using the electric field distribution function $D$. The second method require us to perform 2D-PBE with boundary conditions specified in terms of $\bar{E}$ and $D$.\\
For the first method (method I), we initially compute $\bar{\phi}$ of points $(i,j)$ where index $i$ refers to the $x$ axis and $j$ for the $r$ axis, where the $x$ grid distance are uniform, as with the $r$ grid distances, but they need not be equal in length interval. The details of the solution for $\bar{\phi}$ was detailed in the previous section, where Eq. (\ref{eq:cylPB}) plays a pivotal rule. The solutions are determined for all $(i,j)$ in $(\bar{x},\bar{r})$ space. We then determine a relationship between $\phi$ and $\bar\phi$ (Eq. \ref{eq:kkk}). We then form a grid $(i,j)$ in $(x,r)$ space which is exactly the same as $(\bar{x},\bar{r})$ space. We then compute $\phi$ over the grid $(i,j)$ using our solutions for $\bar{\phi}$, where for solution at grid $(i,j)$ we require $\bar{\phi}$ at \{$(i,j-1)(i,j)(i,j+1)$\}.\\
For method II, we also construct an $(i,j)$ grid over $(x,r)$ space. We then determine the values of $\bar{\phi}$ at the grid boundaries. We also require the derivatives of $\bar{\phi}$ at this boundary which corresponds to the von-Neumann boundary conditions. We then use these boundary condition values to compute the 2-D PBE over a grid in $(x,r)$ space where Eq. (\ref{eq:PB2D}) is solved. Subsection \ref{sub:pot-data} details the consistency of these two approach.
\subsubsection{Method I: Modifying the Potential of a Continuously Charged Rod} \label{sub:Method1}
\begin{figure}[h!]
\begin{pspicture}(-3.2,-1.9)(5,6)
\psset{xunit = 0.85cm, yunit=0.85cm}

\psline[linewidth=2.6pt,linecolor=black!60]{-}(-3.2,0.5)(0.9,0.5)
\psline[linewidth=2.6pt,linecolor=black!60]{-}(-3.2,-0.5)(0.9,-0.5)
\psline[linewidth=1.0pt,linecolor=black!90,linestyle=dotted]{->}(-2.8,0)(1.6,0)
\psline[linewidth=1.0pt,linecolor=black!90,linestyle=dotted]{->}(-2.8,0)(-2.8,6)
\rput(1.4,0.2){$x$}
\rput(-2.95,5.5){$r$}
\rput(-1.2,0.3){$+\,+\,+\,+\,+\,+\,+\,+\,+$}
\rput(-1.2,-0.3){$+\,+\,+\,+\,+\,+\,+\,+\,+$}
\psline[linewidth=3.0pt,linecolor=black!60]{->}(-2.5,0.5)(-2.5,3.2)
\psline[linewidth=3.0pt,linecolor=black!60]{->}(-2.0,0.5)(-2.0,3.2)
\psline[linewidth=3.0pt,linecolor=black!60]{->}(-1.5,0.5)(-1.5,3.2)
\psline[linewidth=3.0pt,linecolor=black!60]{->}(-1.0,0.5)(-1.0,3.2)
\psline[linewidth=3.0pt,linecolor=black!60]{->}(-0.5,0.5)(-0.5,3.2)
\psline[linewidth=3.0pt,linecolor=black!60]{->}(0.0,0.5)(0.0,3.2)
\rput(-1.8,3.7){$\bar{\mathbf{E}}(r)$}
\rput(-0.3,3.7){$\bar{\phi}(r)$}
\rput(-1.1,-1.6){(a)}

\psline[linewidth=0.5pt,linecolor=black!90](5.5,0.5)(5.5,4.5)
\psline[linewidth=0.5pt,linecolor=black!90](6,0.5)(6,4.5)
\psline[linewidth=0.5pt,linecolor=black!90](6.5,0.5)(6.5,4.5)
\psline[linewidth=0.5pt,linecolor=black!90](7,0.5)(7,4.5)
\psline[linewidth=0.5pt,linecolor=black!90](7.5,0.5)(7.5,4.5)
\psline[linewidth=0.5pt,linecolor=black!90](5.0,1)(8.0,1)
\psline[linewidth=0.5pt,linecolor=black!90](5.0,1.5)(8.0,1.5)
\psline[linewidth=0.5pt,linecolor=black!90](5.0,2)(8.0,2)
\psline[linewidth=0.5pt,linecolor=black!90](5.0,2.5)(8.0,2.5)
\psline[linewidth=0.5pt,linecolor=black!90](5.0,3.0)(8.0,3.0)
\psline[linewidth=0.5pt,linecolor=black!90](5.0,3.5)(8.0,3.5)
\psline[linewidth=0.5pt,linecolor=black!90](5.0,4.0)(8.0,4.0)

\psline[linewidth=2.6pt,linecolor=black!60]{-}(4.0,0.5)(9.0,0.5)
\psline[linewidth=2.6pt,linecolor=black!60]{-}(4.0,-0.5)(9.0,-0.5)
\psline[linewidth=1.0pt,linecolor=black!90,linestyle=dotted]{->}(5.5,0)(10.5,0)
\psline[linewidth=1.0pt,linecolor=black!90,linestyle=dotted]{->}(5.5,0)(5.5,6)
\rput(10.0,0.2){$x$}
\rput(5.25,5.5){$r$}
\psdot[dotscale=2.0](5.5,0.0)
\psdot[dotscale=2.0](7.5,0.0)
\rput(4.4,-1.0){rod charge}
\psline[linewidth=0.7pt,linecolor=black!90]{->}(4.6,-0.8)(5.3,-0.2)
\psline[linewidth=0.7pt,linecolor=black!90]{|-}(5.52,-0.9)(6.3, -0.9)
\psline[linewidth=0.7pt,linecolor=black!90]{-|}(6.9,-0.9)(7.5, -0.9)
\rput(6.6,-0.9){$b$}
\rput(8.5,-1.6){(b)}

\pscircle(7.0,3.5){0.1}
\rput(7.2,3.3){$\small c$}
\psdot[dotscale=0.8](7.0,3.0)
\psdot[dotscale=0.8](7.0,4.0)
\psdot[dotscale=0.8](7.5,3.5)
\psdot[dotscale=0.8](6.5,3.5)
\pscircle[linewidth=0.7pt,linecolor=black!60,linestyle=dashed](7.0,3.5){0.68}
\psline[linewidth=1.5pt,linecolor=black!100]{->}(6.6,3.5)(7.4,3.5)
\psline[linewidth=1.5pt,linecolor=black!100]{->}(7,3.1)(7,3.9)

\pscurve[linecolor=black!70]{->}(7.3,4.4)(8.3,5.8)(9.5,6.2)(10.0,6.0)
\pscircle[linewidth=0.7pt,linecolor=black!60,linestyle=dashed](12.5,4.0){2.5}
\psline[linewidth=0.5pt,linecolor=black!90](10.3,2.0)(14.7,2.0)
\psline[linewidth=0.5pt,linecolor=black!90](9.5,4.0)(15.4,4.0)
\psline[linewidth=0.5pt,linecolor=black!90](10.3,6.0)(14.7,6.0)
\psline[linewidth=0.5pt,linecolor=black!90](10.5,1.8)(10.5,6.2)
\psline[linewidth=0.5pt,linecolor=black!90](12.5,1.1)(12.5,6.9)
\psline[linewidth=0.5pt,linecolor=black!90](14.5,1.8)(14.5,6.2)
\psline[linewidth=2.0pt,linecolor=black!100]{->}(11.3,4.0)(13.7,4.0)
\psline[linewidth=2.0pt,linecolor=black!100]{->}(12.5,2.8)(12.5,5.2)
\pscircle(12.5,4.0){0.2}
\rput(12.8,3.7){\normalsize $c$}
\rput(11.9,3.6){\small $(x_i,r_j)$}
\psdot[dotscale=1.5, dotstyle=*](10.5,4.0)
\rput(9.8,3.7){\small $(x_{i-1},r_j)$}
\psdot[dotscale=1.5, dotstyle=*](14.5,4.0)
\rput(15.1,3.7){\small $(x_{i+1},r_j)$}
\psdot[dotscale=1.5, dotstyle=*](12.5,2.0)
\rput(13.2,1.6){\small $(x_i,r_{j-1})$}
\psdot[dotscale=1.5, dotstyle=*](12.5,6.0)
\rput(13.2,6.4){\small $(x_i,r_{j+1})$}
\rput(13.5,4.4){\normalsize $\frac{\partial D_x}{\partial x}$}
\rput(12.1,4.65){$\frac{\partial D_y}{\partial r}$}
\rput(12.1,5.35){$\frac{\partial \bar{\phi}}{\partial r}$}
\psline[linewidth=0.5pt,linecolor=black!90]{<->}(10.5,2.15)(12.5,2.15)
\rput(11.5,2.35){\small $h_x$}
\psline[linewidth=0.5pt,linecolor=black!90]{<->}(10.65,2.0)(10.65,4.0)
\rput(10.9,3.0){\small $h_r$}

\end{pspicture}
\caption{\label{fig:cylinderI}(a). Electric field ($\bar{\mathbf{E}}(r)$) and potential ($\bar{\phi}(r)$) of a continuously charged rod as a function of coordinate $r$ only. (b). An example of a central point $(x_i, r_j)$ and its neighbor points for the $\phi(x_i, r_j)$ potential calculation with method I (Eq. \ref{eq:FD-pot})}
\end{figure}
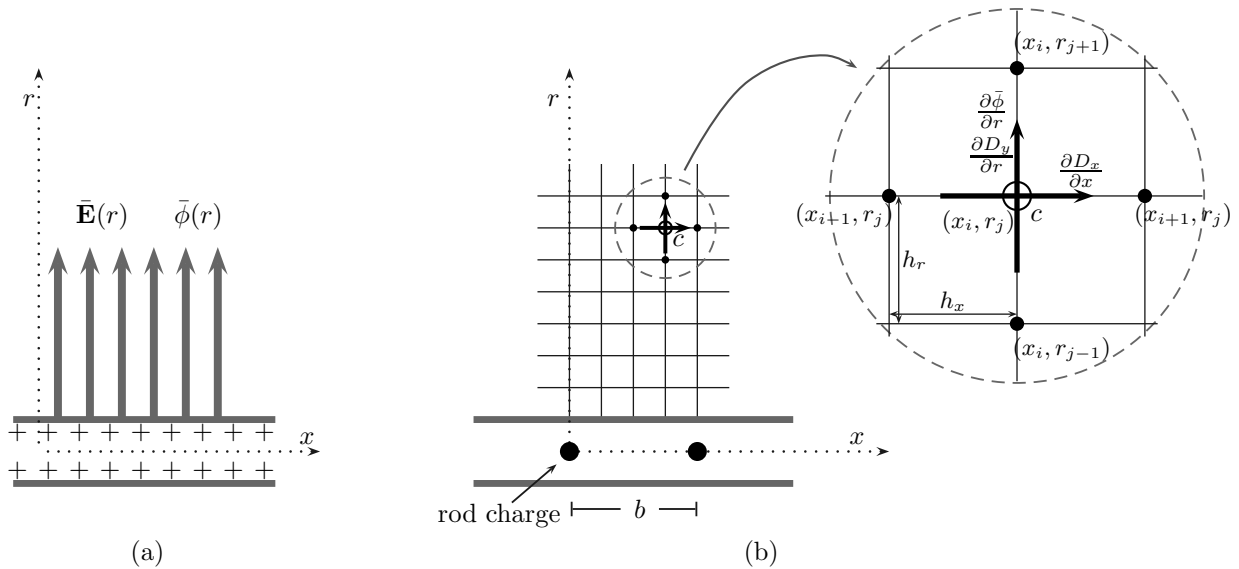

The starting point is the Poisson equation (mks unit)
\begin{align}
\nabla\cdot{\mathbf{E}}(r^\prime) &= \frac{\rho(r^\prime)}{\epsilon} \label{eq:max1}\text{ , }
\end{align}
where $\rho(r^{\prime})$ is the charge density at $r^{\prime}$ and $\epsilon$ dielectric constant. The coordinate ($r^\prime$) where the electric field $\mathbf{E}(r^\prime)$ and charge density $\rho(r^\prime)$ occur are outside the rod volume. We then assume the $\epsilon$ at the medium outside the rod is constant ($r^\prime$ independent). If we denote $\rho$ as the charge density around a discretely charged rod and similarly $\bar{\rho}$ for the continuous charged rod, then
\begin{align}
\frac{\nabla \cdot \mathbf{E}(r^\prime)}{\nabla \cdot \mathbf{\bar{E}}(r^\prime)} &= \frac{\rho(r^\prime)}{\bar{\rho}(r^\prime)} \label{eq:klm}
\end{align}
In cylindrical coordinate
\begin{align}
\nabla \cdot \mathbf{E} & = \frac{1}{r} \left ( \frac{\partial(rE_r)}{\partial r} \right ) + \frac{1}{r} \left ( \frac{\partial E_{\theta}}{\partial \theta} \right ) + \left ( \frac{\partial E_x}{\partial x} \right ) \text{ . } \label{eq:nablaE}
\end{align}
The $y$ axis in Fig. \ref{fig:mon-periodic} is identical with the coordinate $r$ in cylindrical coordinate. We will use the $r$ symbol to represent the coordinate at $\theta=0$ in cylindrical coordinates. Therefore any arbitrary point $r^\prime$ (\ref{eq:max1}) is a function of $r,\,\theta$ and $x$. The electric field and charge density for discrete rod model is ($r,\,x$) dependent and for the continuous model is ($r$) dependent. Then (\ref{eq:klm}) becomes
\begin{equation}
\frac{\frac{1}{r} \left ( \frac{\partial(rE_r)}{\partial r} \right ) + \left ( \frac{\partial E_x}{\partial x} \right )}{\frac{1}{r} \left ( \frac{\partial(r\bar{E})}{\partial r} \right )} = \frac{\rho(x,r)}{\bar{\rho}(r)}\label{eq:aaa}
\end{equation}
Substituting $E_r$ for $\bar{E}_r$ in (\ref{eq:DxE}$\--$\ref{eq:DE}), (\ref{eq:aaa}) becomes
\begin{align}
\frac{\frac{1}{r} \left ( \frac{\partial(rD_y\bar{E})}{\partial r} \right ) + \frac{\partial (D_x\bar{E})}{\partial x}}{\frac{1}{r} \left ( \frac{\partial(r\bar{E})}{\partial r} \right )} & = \frac{\rho(x,r)}{\bar{\rho}(r)}\\
\intertext{which implies}
\frac{D_y\frac{1}{r} \left ( \frac{\partial(r\bar{E})}{\partial r} \right ) + \bar{E} \frac{\partial D_y}{\partial r} + \bar{E}\frac{\partial D_x}{\partial x}}{\frac{1}{r} \left ( \frac{\partial(r\bar{E})}{\partial r} \right )} & = \frac{\rho(x,r)}{\bar{\rho}(r)}\\
\intertext{or}
D_y + \frac{\bar{E}\left( \frac{\partial D_y}{\partial r} + \frac{\partial D_x}{\partial x} \right )}{\frac{\partial\bar{E}}{\partial r} + \frac{1}{r} \bar{E}}  & = \frac{\rho(x,r)}{\bar{\rho}(r)} \label{eq:qwe}
\end{align}
In this work we approximate $\rho(r^{\prime})$ by the Boltzmann factor $\rho_s(r^{\prime}) = \rho_{s,\infty}\,z_se^{-z_s\phi(r^\prime)}$ where $\rho_{s,\infty}$ and $z_s$ are the bulk concentration and finite charge of species $s$. Since $E = -\nabla\phi$ and $\bar{E}$ is only $r$ dependent, then (\ref{eq:qwe}) becomes
\begin{align}
\sum_s \rho_{s,\infty}\,z_se^{-z_s\phi(x,r)} &= \left [ D_y + \frac{\frac{\partial\bar\phi}{\partial r}\left( \frac{\partial D_y}{\partial r} + \frac{\partial D_x}{\partial x} \right )}{\frac{\partial^2\bar\phi}{\partial r^2} + \frac{1}{r}\frac{\partial\bar\phi}{\partial r}} \right ] \sum_i \rho_{s,\infty}\,z_se^{-z_s\bar{\phi}(r)} \text{ .} \label{eq:kkk}
\end{align}
We need the new potential $\phi(x,r)$ in the LHS. We propose to do numerical computation of (\ref{eq:kkk}) by replacing the derivatives with their numerical differencing form \cite{Fox57}. For a second order numerical differencing, one possibility for (\ref{eq:kkk}) is 
\begin{align}
\sum_s \rho_{s,\infty}\,z_se^{-z_s\phi(x_i,r_j)} =& \left [D_y(x_i,r_j) + \frac{\frac{\bar{\phi}(x_i,r_{j+1})-\bar{\phi}(x_i,r_{j-1})}{2h_r} \left (\frac{D_y(x_i,r_{j+1}) - D_y(x_i,r_{j-1})}{2h_r} + \frac{D_x(x_{i+1},r_j)-D_x(x_{i-1},r_j)}{2h_x} \right ) }{\frac{\bar{\phi}(x_i,r_{j+1}) - 2\bar{\phi}(x_i,r_j) + \bar{\phi}(x_i,r_{j-1})}{h_r^2} + \frac{1}{r}\frac{\bar{\phi}(x_i,r_{j+1})-\bar{\phi}(x_i,r_{j-1})}{2h_r}} \right ]\notag{}\\
 &\times \sum_s \rho_{s,\infty}\,z_se^{-z_s\bar{\phi}(x_i,r_j)} \text{ .} \label{eq:FD-pot}
\end{align}
For reasons that follow, only a 1-D root finding technique is required to solve (\ref{eq:FD-pot}). Fig.\ref{fig:cylinderI}.b illustrates a sample point $(x_i, r_j)$ and its four nearest neighboring points $\left [ (x_{i+1}, r_j),(x_{i-1}, r_j),(x_i, r_{j+1}),(x_i, r_{j-1}) \right ]$. By supplying the $\bar{\phi}$ and $D$ values of the sample point $(x_i, r_j)$ and its neighboring point to the RHS of (\ref{eq:FD-pot}), the RHS of (\ref{eq:FD-pot}) can be computed. We then determine the potential $\phi(x_i,r_j)$ for in the LHS by the method of root finding, i.e bisection or Newton-Raphson method. For the computation at the charged rod-solution interfaces, we can choose either the boundary conditions that obtain at the interface or reduce the order of numerical differencing into its first order. We use the boundary conditions specified in Subsection \ref{subsec:MultPBE} for $\bar{E}$ values at boundaries.\\
We close this subsection by giving an additional note on (\ref{eq:kkk}). From Fig. \ref{fig:normal-Dy} and \ref{fig:normal-Dx}, when the perpendicular distance $r$ to the rod surface is far enough, the value of $D_y$ is constant unity $(\,=1)$ and the value of $D_x$ is constant zero $(\,=0)$. It makes the term $\frac{\partial{D_y}}{\partial{r}}$ and $\frac{\partial{D_x}}{\partial{x}}$ in the square bracket of (\ref{eq:kkk}) equal zero. Then total value inside the square bracket of (\ref{eq:kkk}) equals $D_y=1$ leading to $ \phi(x,r) = \bar{\phi}(r)$. Thus at large distance, the potential around a discretely charged rod is equal to its continuous counterpart.\\
\subsubsection{Method II: Applying the 2-Dimensional PBE} \label{sub:Method2}
We propose another method to obtain the non-isotropic potential by including the electric field distribution function $D$ in the numerical potential calculation of PBE.\\ 
\begin{figure}[h!]
\begin{pspicture}(-4,-1.5)(6,6)
\psset{Alpha=75, Beta=15, xunit=1.3cm, yunit=1.3cm}
\parametricplotThreeD[xPlotpoints=100,linecolor=black,linewidth=1.5pt,plotstyle=curve](0,360){1.5 t cos mul 1.5 add 1.5 t sin mul 1.5 add 0 }
\parametricplotThreeD[xPlotpoints=10,linecolor=black,linewidth=1.5pt](0,4){1.5 0 1 t mul }
\parametricplotThreeD[xPlotpoints=10,linecolor=black,linewidth=1.5pt](0,4){1.5 3 1 t mul }
\parametricplotThreeD[xPlotpoints=10,linecolor=black!50,linewidth=20pt](0.0,4.0){1.5 1.5 1 t mul }
\parametricplotThreeD*[xPlotpoints=100,linecolor=black!50,linewidth=1.0pt,plotstyle=curve](0,360){0.25 t cos mul 1.5 add 0.25 t sin mul 1.5 add 0 }
\parametricplotThreeD[xPlotpoints=10,linecolor=black!50,linewidth=20pt](2.6,4){1.5 1.5 1 t mul }
\pstThreeDPut[linecolor=black!90](1.7,1.6,0.58){\Large{$\--$}}
\pstThreeDPut[linecolor=black!90](1.7,1.6,0.57){\Large{$\--$}}
\pstThreeDPut[linecolor=black!90](1.7,1.6,2.40){\Large{$\--$}}
\pstThreeDPut[linecolor=black!90](1.7,1.6,2.38){\Large{$\--$}}
\pstThreeDPut[linecolor=black!90](1.7,1.6,4.17){\Large{$\--$}}
\pstThreeDPut[linecolor=black!90](1.7,1.6,4.18){\Large{$\--$}}

\parametricplotThreeD[xPlotpoints=100,linecolor=black,linewidth=1.5pt,plotstyle=curve](0,360){1.5 t cos mul 1.5 add 1.5 t sin mul 1.5 add 4 }
\parametricplotThreeD[xPlotpoints=10,linecolor=black!50,linewidth=20pt](4.0,4.7){1.5 1.5 1 t mul }
\parametricplotThreeD[xPlotpoints=100,linecolor=black!50,linewidth=1.0pt,plotstyle=curve](0,360){0.25 t cos mul 1.5 add 0.25 t sin mul 1.5 add 4.7 }
\parametricplotThreeD*[xPlotpoints=100,linecolor=black!50,linewidth=1.0pt,plotstyle=curve](0,360){0.25 t cos mul 1.5 add 0.25 t sin mul 1.5 add 2.6}

\psline[linewidth=1.0pt](1.22,0.03)(2.5,-0.08)
\psline[linewidth=1.0pt](1.22,1.0)(2.5,0.88)
\psline[linewidth=1.0pt](1.22,0.03)(1.22,1.0)
\psline[linewidth=1.0pt,linecolor=black!60](1.3,0.5)(1.4,0.9)
\psline[linewidth=1.0pt,linecolor=black!60](1.3,0.12)(1.58,0.9)
\psline[linewidth=1.0pt,linecolor=black!60](1.5,0.10)(1.8,0.88)
\psline[linewidth=1.0pt,linecolor=black!60](1.7,0.08)(2.0,0.86)
\psline[linewidth=1.0pt,linecolor=black!60](1.9,0.06)(2.2,0.84)
\psline[linewidth=1.0pt,linecolor=black!60](2.1,0.05)(2.4,0.82)
\psline[linewidth=1.0pt,linecolor=black!60](2.3,0.05)(2.48,0.45)
\rput(1.2,-0.07){\bf{A}}
\rput(1.1,1.0){\bf{B}}
\rput(2.65,0.95){\bf{C}}
\rput(2.65,-0.03){\bf{D}}
\psline[linewidth=0.8pt]{<->}(1.08,2.2)(2.5,2.1)
\rput(2.0, 2.3){$\sub{r}{c}$}
\psline[linewidth=0.7pt]{<->}(1.05,2.65)(1.37,2.62)
\rput(1.6, 2.6){$\sub{r}{p}$}

\psline[linewidth=0.7pt]{|-}(0.6,0.01)(0.6,0.7)
\psline[linewidth=0.7pt]{-|}(0.6,1.2)(0.6,1.8)
\rput(0.6, 0.9){$b$}

\psline[linewidth=0.6pt,linecolor=black!60]{->}(1.05,0.06)(1.05,4.6)
\rput(0.9, 4.5){$\mathrm{x}$}
\psline[linewidth=0.6pt,linecolor=black!60]{->}(1.05,0.06)(3.3,-0.2)
\rput(3.3, 0.0){$\mathrm{r}$}
\psline[linewidth=0.6pt,linecolor=black!60]{->}(1.05,0.06)(-0.3,-0.85)
\rput(-0.1, -1.0){$\mathrm{z}$}
\psline[linewidth=0.6pt,linecolor=black!60, linestyle=dashed](1.05,3.5)(2.55,3.35)
\psline[linewidth=1.0pt,linecolor=black!90]{->}(2.55,3.35)(2.8,3.55)
\psline[linewidth=0.6pt,linecolor=black!60, linestyle=dashed](1.05,3.5)(0.0,3.1)
\pscurve[linecolor=black!90,linewidth=0.4pt](0.6,3.32)(0.85,3.24)(1.1,3.22)(1.3,3.25)(1.56,3.4)(1.6,3.45)
\rput(1.23, 3.35){$\theta$}
\rput(2.95, 3.45){$\mathrm{\theta}$}

\pspolygon[linewidth=2pt,linecolor=black!90](5.5,0.5)(7.5,0.5)(7.5,3.5)(5.5,3.5)
\psline[linewidth=1pt,linecolor=black!90](6,0.5)(6,3.5)
\psline[linewidth=1pt,linecolor=black!90](6.5,0.5)(6.5,3.5)
\psline[linewidth=1pt,linecolor=black!90](7,0.5)(7,3.5)
\psline[linewidth=1pt,linecolor=black!90](5.5,1)(7.5,1)
\psline[linewidth=1pt,linecolor=black!90](5.5,1.5)(7.5,1.5)
\psline[linewidth=1pt,linecolor=black!90](5.5,2)(7.5,2)
\psline[linewidth=1pt,linecolor=black!90](5.5,2.5)(7.5,2.5)
\psline[linewidth=1pt,linecolor=black!90](5.5,3.0)(7.5,3.0)

\psline[linewidth=0.6pt,linecolor=black!60]{->}(5.5,-1)(5.5,4.5)
\psline[linewidth=0.6pt,linecolor=black!60]{->}(4.5,0)(8.5,0)
\rput(8.5,0.3){$x$}
\rput(5.25,4.2){$r$}
\psdot[dotscale=2.0](5.5,0.0)
\rput(5.3,0.4){$\mathbf{A}$}
\rput(7.7,0.4){$\mathbf{B}$}
\rput(7.7,3.5){$\mathbf{C}$}
\rput(5.3,3.5){$\mathbf{D}$}
\rput(4.6,-0.8){rod charge}
\psline[linewidth=0.7pt,linecolor=black!90]{->}(4.8,-0.6)(5.3,-0.2)
\psline[linewidth=0.7pt,linecolor=black!90]{|-}(5.52,-0.7)(6.3, -0.7)
\psline[linewidth=0.7pt,linecolor=black!90]{-|}(6.9,-0.7)(7.5, -0.7)
\rput(6.55,-0.7){$\frac{1}{2}b$}
\psline[linewidth=0.7pt,linecolor=black!90]{|-}(8,0.52)(8, 1.7)
\psline[linewidth=0.7pt,linecolor=black!90]{-|}(8,2.3)(8, 3.5)
\rput(8,2.0){$\sub{r}{c} - \sub{r}{p}$}
\psline[linewidth=0.7pt,linecolor=black!90]{|-}(4.85,0.0)(4.85, 1.5)
\psline[linewidth=0.7pt,linecolor=black!90]{-|}(4.85,2.0)(4.85, 3.5)
\rput(4.85,1.7){$\sub{r}{c}$}
\end{pspicture}
\caption{\label{fig:cylinderII}$\mathbf{ABCB}$ plane. Left: Point $\mathbf{A}$ lies at a distance $\sub{r}{p}$ which is exactly perpendicular to a rod charge. Point $\mathbf{B}$ is perpendicular to the middle of two adjacent charges, point $\mathbf{C}$ and $\mathbf{D}$ lies at the cylinder box. Right: The plane $\mathbf{ABCD}$ is divided into grids for numerical computation.}
\end{figure}
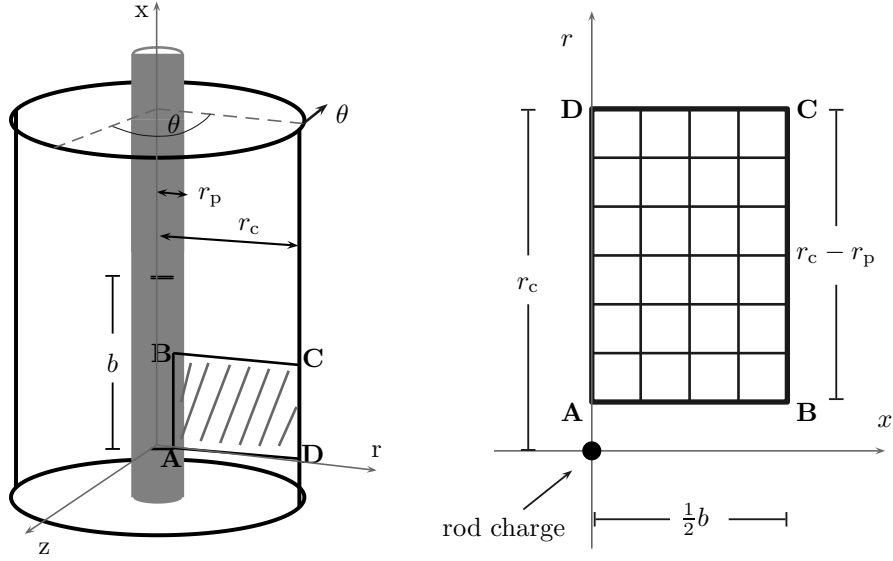
By (\ref{eq:max1}) and (\ref{eq:nablaE}) we have 
\begin{align}
\frac{1}{r} \left ( \frac{\partial(rE_r)}{\partial r} \right ) + \frac{1}{r} \left ( \frac{\partial E_{\theta}}{\partial \theta} \right ) + \left ( \frac{\partial E_x}{\partial x} \right ) &= \frac{\rho(r,\theta,x)}{\epsilon} \text{ . }
\intertext{The electric field over $\theta$ coordinate is isotropic (see Fig. \ref{fig:cylinderII}), thus we can omit the $\theta$ terms. Then} 
\left ( \frac{\partial E_r}{\partial r} \right ) + \frac{1}{r} E_r + \left ( \frac{\partial E_x}{\partial x} \right ) &= \frac{\rho(r,x)}{\epsilon} \label{eq:E2D}\text{ . }
\end{align}
Recall that \citep{Sch87}
\begin{equation}
\mathbf{E} = -\nabla{\phi} = - \left (\frac{\partial\phi}{\partial r} \right ) \hat{\mathbf{r}} - \frac{1}{r} \left( \frac{\partial\phi}{\partial\theta} \right ) \hat{\mathbf{\theta}} - \left ( \frac{\partial \phi}{\partial x} \right ) \hat{\mathbf{x}} \notag{}
\end{equation}
Then by expressing the charge density $\rho(r,x)$ by the Boltzmann factor, (\ref{eq:E2D}) becomes
\begin{equation}
\left ( \frac{\partial^2 \phi}{\partial r^2} \right ) + \frac{1}{r} \frac{\partial\phi}{\partial r} + \left ( \frac{\partial^2 \phi}{\partial x^2} \right ) = -\sum_i^s \frac{e^2z_i\rho_\infty^i \exp(-z_i\phi)}{\epsilon \sub{k}{B}T} \text{ , } \label{eq:PB2D}
\end{equation}
where $\rho_\infty^i$ and $z_i$ are the bulk density and finite charge of species $i$; $s$ is the total species in the system except for the charged cylinder.\\
Eq. (\ref{eq:PB2D}) is a partial differential equation with independent variables $r$ and $x$. In the $x$ direction, the integration is performed over $x\in[0,1/2b]$ because $\phi(r,x)$ is symmetric and periodic. We define a small rectangular plane $\mathbf{ABCD}$ (Fig \ref{fig:cylinderII}) with plane width $\mathbf{AB}=\mathbf{CD}=1/2b$. The plane length $\mathbf{BC}=\mathbf{AD}$ = $\sub{r}{c} - \sub{r}{p}$, where $\sub{r}{c}$ is the cylindrical box radius and $\sub{r}{p}$ is the closest contact distance between the rod axis and the counterion center. The corner point $\mathbf A$ is set to lie exactly perpendicular to the discrete rod charge with distance $\sub{r}{p}$ from the rod axis. Point $B$ lies exactly on the perpendicular to the rod axis, pass through the mid-point of adjacent rod charges with distance $\sub{r}{p}$ from the rod axis. To solve the potential solution within the $\mathbf{ABCD}$, we specify the Neumann boundary conditions at the edges of the $\mathbf{ABCD}$ plane. The potential derivatives which is the negative of the electric fields are specified by combining the electric field distribution function $D$ and the electric fields of the continuous charged rod. Then the boundary conditions for (\ref{eq:PB2D}) at the $\mathbf{ABCD}$ plane are
\begin{align}
\left (\frac{\partial \phi}{\partial r} \right )_\mathbf{AB} &= -D_y(\sub{r}{p},x)\,\bar{E}(\sub{r}{p}) &\qquad \left (\frac{\partial \phi}{\partial r} \right )_\mathbf{BC} &= -D_y(r,\frac{1}{2}b)\,\bar{E}(r) \label{eq:Bound-1}\\
\left (\frac{\partial \phi}{\partial x} \right )_\mathbf{AB} &= -D_x(\sub{r}{p},x)\,\bar{E}(\sub{r}{p}) &\qquad \left (\frac{\partial \phi}{\partial x} \right )_\mathbf{BC} &= 0\\
\left (\frac{\partial \phi}{\partial r} \right )_\mathbf{CD} &= 0 &\qquad \left (\frac{\partial \phi}{\partial r} \right )_\mathbf{AD} &= -D_y(r,0)\,\bar{E}(r) \\
\left (\frac{\partial \phi}{\partial x} \right )_\mathbf{CD} &= -D_x(\sub{r}{c},x)\,\bar{E}(\sub{r}{c}) = 0 &\qquad \left (\frac{\partial \phi}{\partial x} \right )_\mathbf{AD} &= 0\label{eq:Bound-4}
\end{align}
The electric field in the $x$ direction of the boundary $\mathbf{BC}$ and $\mathbf{AD}$ equals zero because of electric field cancellation due to rod charge symmetry. $\bar{E}(r)$ is the electric field of a continuous charged rod model having the same dimension and line charge density as with the discretely charges rod. One possibility to obtain $\bar{E}(r)$ is to calculate the potential gradient of $\bar\phi(r)$ by numerical differencing. Thus before we find the potential solution of a discretely charged rod, first we need to obtain the potential of the equivalent continuously charged rod to define the boundary conditions.\\
\section{Calculating the Mean Radial Distribution Function (RDF) from the Potential Data} \label{sec:get-rdf}
At this point we assume that we have been able to obtain the potential at any point around a charged rod. We then compare a potential-related property, which is the radial distribution function, obtained from simulation and calculation. Radial distribution function $g_{s,s_1}(r^\prime)= \tilde{\rho}_{s,s_1}(r^\prime)/\rho_{s,\infty}$ measures ratio of the average to bulk density of particle $s$ at radial distance $r^\prime$ from the central particle $s_1$. In this article, the central particle is the rod charge and the surrounding particles are the counterion and salt. As such we suppress the $s_1$ subscript and only retain the $s$ label.\\
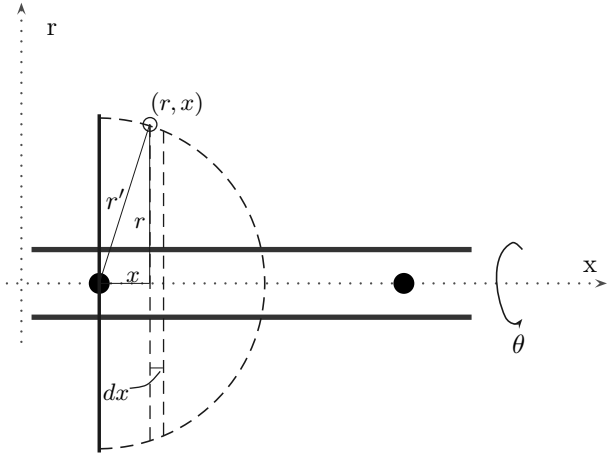
\begin{figure}[h!]
\begin{pspicture}(-2.5,-1)(3,5.7)
\psset{xunit=0.45cm, yunit=0.45cm}

\psline[linecolor=black!70,linewidth=0.9pt, linestyle=dotted]{->}(-3.0,4.0)(15,4.0)
\rput(14.5,4.5){$\mathrm{x}$}
\pscurve[linecolor=black!90, linewidth=0.6pt]{->}(12.5,5.0)(12.2,5.2)(12.0,3.0)(12.3,2.8)(12.5,2.95)
\rput(12.4,2.2){$\mathrm{\theta}$}
\psline[linecolor=black!80,linewidth=2pt]{-}(-2,5.0)(11,5.0)
\psline[linecolor=black!80,linewidth=2pt]{-}(-2,3.0)(11,3.0)
\psdot[dotscale=2.2](0,4)
\psdot[dotscale=2.2](9,4)

\psline[linecolor=black!70,linewidth=0.9pt, linestyle=dotted]{->}(-2.3,2.0)(-2.3,12.3)
\rput(-1.4,11.5){$\mathrm{r}$}
\psline[linecolor=black!90,linewidth=0.5pt, linestyle=dashed](1.5,-0.65)(1.5,8.7)
\psline[linecolor=black!90,linewidth=0.5pt, linestyle=dashed](1.9,-0.5)(1.9,8.5)

\psline[linecolor=black!90,linewidth=0.15pt](0,4)(1.49,8.7)
\psline[linecolor=black!90,linewidth=0.15pt](0,4)(1.49,4)
\psline[linecolor=black!90,linewidth=0.15pt](1.49,4)(1.49,8.7)
\psline[linecolor=black!90,linewidth=0.4pt](1.5,1.5)(1.88,1.5)
\rput(0.5,0.8){\small{$dx$}}
\pscurve[linecolor=black!90, linewidth=0.5pt](0.9,0.9)(1.4,1.0)(1.8,1.4)
\rput(0.5,6.5){$r^\prime$}
\rput(1.2,5.8){\small{$r$}}
\rput(1.0,4.2){\small{$x$}}
\pscircle[linecolor=black!90,linewidth=0.5pt](1.5,8.7){0.1}
\rput(2.3,9.3){\small{$(r,x)$}}

\psarc[linecolor=black!90,linewidth=0.6pt, linestyle= dashed](0,4){2.2}{-90}{90}
\psline[linecolor=black!90,linewidth=1.3pt](0,-1)(0,9)
\end{pspicture}
\caption{\label{fig:rdf-sph}Radial distance $r^\prime$ of a finite rod charge (closed circle). The rod axis is in the $x$ direction.}
\end{figure} 
Since the potential from the cylindrical PBE is a function of the perpendicular distance $r$ to the rod axis and the lateral $x$ distance from a fixed reference charge on the rod, we need a special treatment to obtain the mean RDF around a rod charge.\\
Fig. \ref{fig:rdf-sph} depicts a dashed-line hemisphere with radius $r^\prime$ and surface area $A$ centered at a rod charge. The average number density at the hemisphere surface is
\begin{align}
\tilde{\rho}_s(r^\prime) &= \frac{\int_{S} ~\rho_s(r,x,\theta) \mathrm{d}A}{A} \text{ , } \label{eq:rho-sph}
\intertext{where $\rho_s(r,x,\theta)$ is the number density at any point at the surface. The density is isotropic over $\theta$ coordinate, thus we can exclude the $\theta$ dependency. The surface area integration of (\ref{eq:rho-sph}) follows the formula for a sphere. (\ref{eq:rho-sph}) then becomes}
\tilde{\rho}_s(r^\prime) &= \frac{2 \pi r^\prime \int_0^{r^\prime} \rho_s(r,x) \mathrm{d}x}{2\pi (r^\prime)^2} \label{eq:ldd} \text{ . }
\end{align}
However, there is one difference between the calculation of bulk concentration of ions in PBE system and simulation. In PBE, the rod volume is not a space that is accessible to ions, thus the bulk density of ions is calculated by
\begin{equation}  
\superx{\rho_{s,\infty}}{PBE} = \frac{N_s}{\sub{V}{Box} - \sub{V}{Rod}} \text{~, }\label{eq:rho-PBE}
\end{equation}
where $N_s$ is the total number of ion $s$, $\sub{V}{Box}$ is the box volume and $\sub{V}{Rod}$ is the rod volume.\\
In simulation, the calculation of the ionic bulk density involves the whole box volume. So the bulk density of ion $s$ is
\begin{equation}  
\super{\rho_{s,\infty}}{Sim} = \frac{N_s}{\sub{V}{Box}} \text{~, }\label{eq:rho-Sim}
\end{equation}
where the definition of $N_s$ and $\sub{V}{Box}$ are similar with (\ref{eq:rho-PBE}).\\
The $\tilde{\rho}_s(r^\prime)$ in (\ref{eq:ldd}) is the source of the theoretical $g(r^\prime)$ that will be compared to the simulation $g(r^\prime)$. Thus
\begin{equation}  
\tilde{\rho}_s(r^\prime) = \super{\rho_{s,\infty}}{Sim}\,g(r^\prime) \text{~.} \label{eq:qwm}
\end{equation}
0n the other hand, the local density $\rho_s(r,x)$ in (\ref{eq:ldd}) is the data from PBE calculation, so that
\begin{align}
\rho_s(r,x)  &= \superx{\rho_{s,\infty}}{PBE} \, \sub{g}{f}(r,x) \notag{}\\
 & =  \superx{\rho_{s,\infty}}{PBE} \, \exp(-z_s \phi(r,x)) \text{~,} \label{eq:qqq}
\end{align}
where $g_f$ has been defined for the PBE in the introduction. Substituting $\tilde{\rho}_s(r^\prime)$ and $\rho_s(r,x)$ in (\ref{eq:qwm}) and (\ref{eq:qqq}) to (\ref{eq:ldd}), we then have
\begin{equation}  
\super{\rho_{s,\infty}}{Sim}\,g(r^\prime)  = \frac{1}{r^\prime} \int_0^{r^\prime} \superx{\rho_{s,\infty}}{PBE} \, \exp(-z_s \phi(r,x)) \mathrm{d}x \label{eq:rkt}
\end{equation}
Since $\superx{\rho_{s,\infty}}{PBE}$ and $\super{\rho_{s,\infty}}{Sim}$ are constant, (\ref{eq:rkt}) reduces to
\begin{equation}  
g(r^\prime)  = \frac{1}{r^\prime} \int_0^{r^\prime} \exp(-z_s \phi(r,x)) \mathrm{d}x~ \frac{\superx{\rho_{s,\infty}}{PBE}}{\super{\rho_{s,\infty}}{Sim}} \text{~.} \label{eq:lmg}
\end{equation}
From (\ref{eq:rho-PBE}) and (\ref{eq:rho-Sim}), since the total number of particle $N_s$ is conserved in our scheme, we have $\superx{\rho_{s,\infty}}{PBE}\,(\sub{V}{Box} - \sub{V}{Rod}) = \super{\rho_{s,\infty}}{Sim}\,\sub{V}{Box}$. So (\ref{eq:lmg}) becomes
\begin{equation}  
g(r^\prime)  = \frac{1}{r^\prime} \int_0^{r^\prime} \exp(-z_s \phi(r,x)) \mathrm{d}x~ \frac{\sub{V}{Box}}{(\sub{V}{Box} - \sub{V}{Rod})} \text{~.} \label{eq:mean-rdf}
\end{equation}
If the rod volume is much smaller than the box volume $(\sub{V}{Rod} \ll \sub{V}{Box})$, the term $\sub{V}{Box}/(\sub{V}{Box} - \sub{V}{Rod}) \approx 1$\\
Numerical integration of (\ref{eq:mean-rdf}) needs to be performed because there is no closed expression of $\phi(r,x)$ which is numerically computed from (\ref{eq:kkk}) and (\ref{eq:PB2D}). We have used both in this presentation (i.e. Method I and Method II). The potential at any point $(r,x)$ which is not exactly at one of the PBE solution points can be determined by interpolation. The interpolation method that is used in this work is either the polynomial or bicubic interpolation \citep{Pre92}.\\
\section{Simulation Part}\label{sec:sim} 
We perform MD simulations for systems containing the rod model A and B (See Fig. \ref{fig:rod-models}) in different salt concentrations inside a cubic box. The rod dimension in the simulation box is made possible by making a cylindrical wall constraint. We fix the $x$ and $y$ of the rod axis in the middle of the simulation box ($\sub{L}{Box}/2,\sub{L}{Box}/2$), where $\sub{L}{Box}$ is the box length. The rod sides are extended in the $z$ direction and the periodic boundary condition applied will make both ends of the rod at infinity. We set the simulation box length to $408\,\angstrom$ so that if the simulation is for rod model A, the number of discrete charges for the rod in the primary simulation cell (box) is $408/20.4=20$. For the simulations of rod model B, the number of discrete charges for the rod per simulation box is $408/1.7=240$. For both models, the total rod charges inside one simulation box is $-240$. Salt ions are modeled by spheres with radius $2\,\angstrom$ where the charge is $+1$ for Na$^+$ and $-1$ for Cl$^-$ \citep{Sim96,Faw96}. The counterions are chosen to have equal charge and radial dimensions as for Na$^+$ from the salt. We use the Langevin thermostat to maintain the system temperature at $300\,K$ for the whole MD run. The Bjerrum length $7.13$ is used as a parameter representing the continuum water solvent dielectric at this temperature.\\
The purely repulsive Lennard-Jones potential is applied to control the short-range repulsive potential. We set the sum of the soft sphere radius ($\sub{\sigma}{LJ}$) for the LJ interaction between the rod surface and ion to $0.1\,\angstrom$ implying that the sum of the hard sphere radii equals $11.9\,\angstrom$. The rod-ion soft sphere interaction radii is chosen to be very small to make the characteristic contact between ions and the rod surface as close as possible to the PBE boundary potential conditions. In PBE, the counterion cannot approach the rod axis at a distance that is smaller than the the sum of their effective radii ($12\,\angstrom$). We use the ESPResSo package to run the MD simulations \cite{Lim06}. The RDF data are generated by the usual dumping procedure after the system attains equilibrium.\\
\section{Result and Discussion}
\begin{figure}[h!]
\includegraphics[scale=0.8]{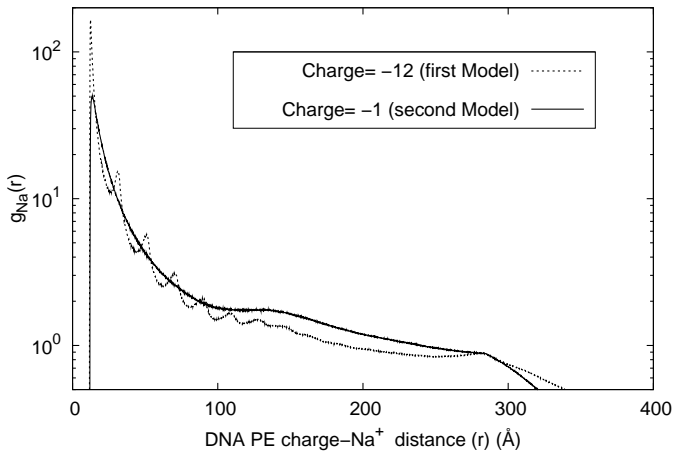}
\caption{\label{fig:rdf-DNA}The RDF profile between the DNA polyelectrolyte charge and Na$^+$ ions. The DNA monomer has charge $-12$ (dotted line) and $-1$ (solid line). The details are discussed in the introduction.}
\end{figure}

The purpose of this work is to examine the consequence of modelling a bunch of charges as one single charge. This is frequently carried out in simulation studies that are interested in the physical forces that determine the particle distribution in complex systems \citep{Hri00}. In these studies, there is an underlying assumption that the simplification of molecular charge distribution in their simulations has a minimal, if not insignificant effect on the resulting general physical properties and distributions of the system. We explicitly show here that this tacit assumption needs to be reconsidered on the basis of Fig. \ref{fig:rdf-DNA}.\\
In order to clarify this situation, we present first our numerical methodology, and will use these techniques in subsequent writings where the PBE will be solved for a DNA polyelectrolyte with different monomer charge and size choices. Then, using our methodology, we can derive the RDF's theoretically and compare it with our simulation results. However, before this can be done, we need to ensure consistency of our method for some model systems, the most amenable being for the continuous and discretely charged rods. We therefore study in depth these systems in what follows. It may be added that this particular approach of the solution to this very important problem in biophysics concerning charge distribution profiles by using a continuously charged reference has not been attempted. The consistency of the result would imply that it is indeed a feasible methodology for all problems related to temperature-dependent charge density distributions.\\     
An example is when 12 (in number) independent charges in the DNA chain is simplified to a single charge, where the chain dimension is conserved.  We had mentioned in the introduction concerning Fig. \ref{fig:rdf-DNA}, where simulation of a linked spherical monomer representing the DNA chain results in the RDF being different when the monomer has a discrete charge $-12$ and $-1$ with the chain dimension and line charge density being conserved for both chains. Here, instead of working with a linked monomer to simulate an array of charge, we use the simpler model which is the infinitely charged rod for rod model A and B. The dimension of the rods are illustrated in  Fig. \ref{fig:rod-models} (for model A and B) where the rod line charge density are the same for the PEs whose RDFs are depicted in Fig. \ref{fig:rdf-DNA}.\\
In what follows we present the potential around the discretely charged rods (for model A and B), as a results of the theoretical calculations utilizing method I and method II (subsection \ref{sub:Method1} and \ref{sub:Method1}). From the potential data, we then construct the theoretical RDF by using the procedure given in Section \ref{sec:get-rdf}. The theoretical RDFs are then compared to the RDFs from MD simulation. The theoretical potential, theoretical RDF, and simulation RDF are studied for both rod models A and B.
\subsection{Potential Data}\label{sub:pot-data}
The equipotential contour lines around the charged rod from the calculations of method I and method II are provided (Fig. \ref{fig:potI} and \ref{fig:potII}). From our numerical data, the potential profiles generated by method I and II are quantitatively the same within computational error suggesting that both methods are equivalent in the potential calculation. Method I is computational cheaper than method II because method I only performs a root-finding procedure at one particular point to obtain the potential at that point, whereas method II numerically calculates the 2-D PBE simultaneously for all the points within the rectangular grid. It may be added that the boundary conditions in method II gives some insight regarding the potential profile within the boundary whereas method I cannot provide such insight.\\
\subsubsection{Method I}
\begin{figure}[h!]
\hspace{0.1cm}
\begin{minipage}[b]{7cm}
\includegraphics[scale=0.75]{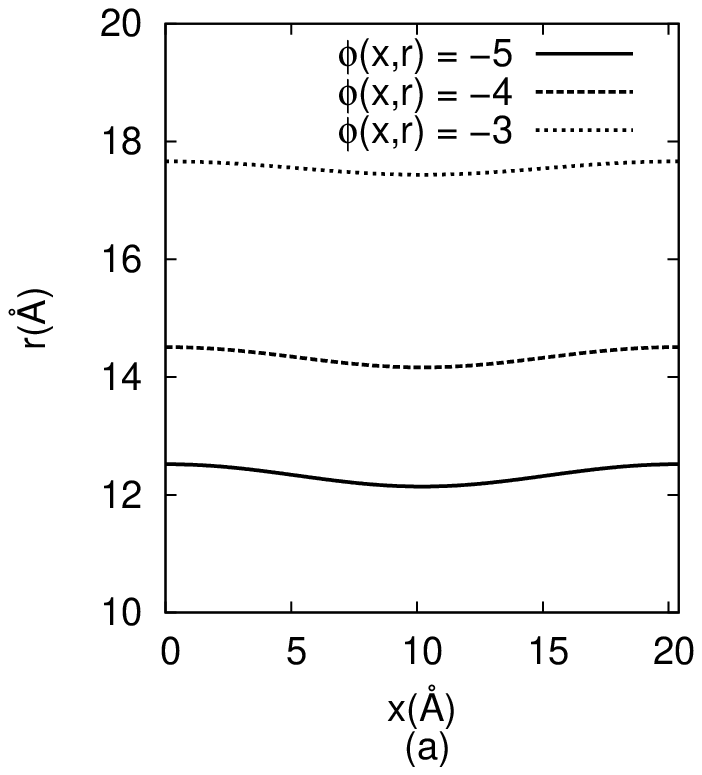}
\end{minipage}
\hspace{0.0cm}
\begin{minipage}[b]{7cm}
\includegraphics[scale=0.75]{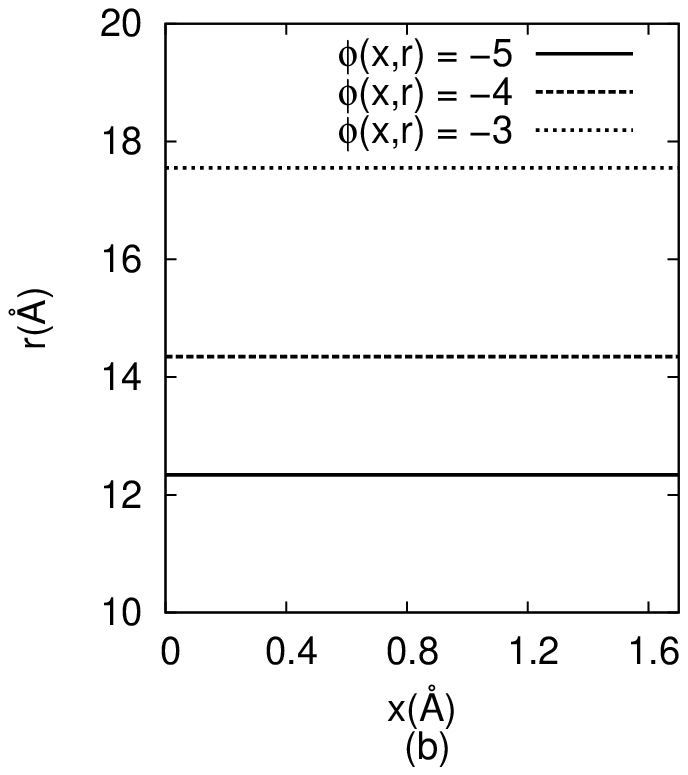}
\end{minipage}
\caption{\label{fig:potI}Potential contour at a distance $r$ perpendicular to a charged rod calculated by method I. Figure (a) for model A ($\subx{z}{P}=-12$, $b=20.4\,\angstrom$) and Fig. (b) for model B ($\subx{z}{P}=-1$, $b=1.7\,\angstrom$) for NaCl salt concentration 10 mM.}
\end{figure}
Figure \ref{fig:potI} shows the $\phi(x,r)$ potential contour profile at a plane whose normal is perpendicular to the rod axis. The fig. \ref{fig:potI}.a for model A, and Fig \ref{fig:potI}.b for model B. The $r$ axis is the perpendicular distance to the rod axis and the $x$ axis is parallel to the rod axis. The range of $x$ axis in Fig \ref{fig:potI} equals the charge-charge distance $b$ for each model. The two adjacent charges are at $(x,r)=(0,0)$ and $(x,r)=(b,0)$. The potential around rod model A varies and is symmetric. The largest absolute value of the potential around model A is at the nearest distance of approach of an ion to the discrete rod charge (at 12 $\angstrom$). In our simulation, the test charge has radius $2\,\angstrom$ and the discrete charges of the rod all have radius $10\,\angstrom$ leading to $12\,\angstrom$ for the distance of closest approach. The potential differences disappear as the distance to the rod surface increases. For model B, there is no significant variation of the potential around the rod axis as $x$ varies.\\
\subsubsection{Method II}
\begin{figure}[h!]
\hspace{-0.2cm}
\begin{minipage}[b]{4.1cm}
\includegraphics[scale=0.85]{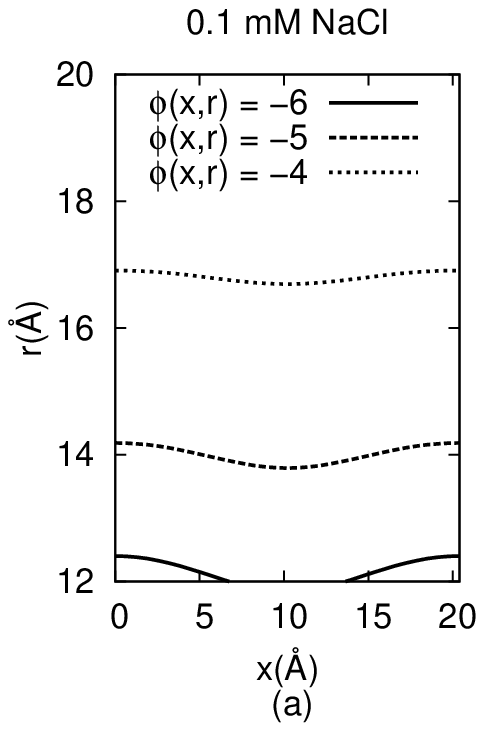}
\end{minipage}
\hspace{-0.1cm}
\begin{minipage}[b]{4.1cm}
\includegraphics[scale=0.85]{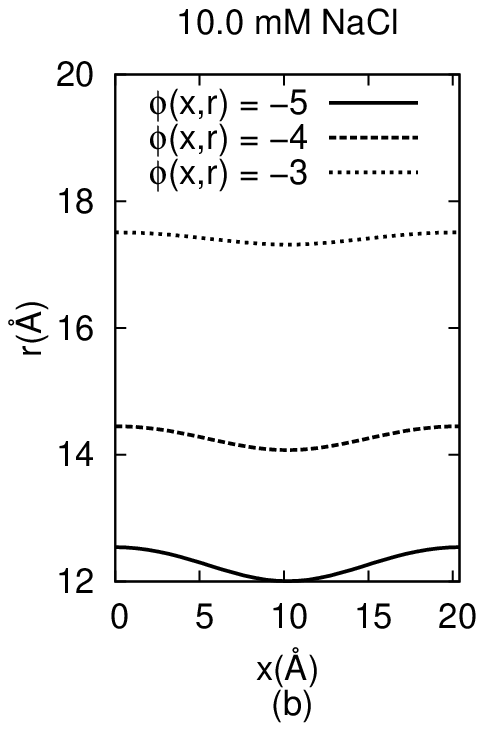}
\end{minipage}
\hspace{-0.1cm}
\begin{minipage}[b]{4.1cm}
\includegraphics[scale=0.85]{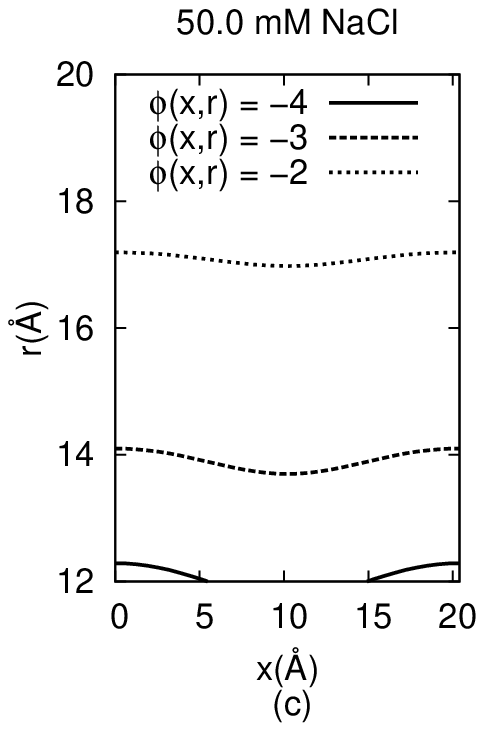}
\end{minipage}
\hspace{-0.1cm}
\begin{minipage}[b]{4.1cm}
\includegraphics[scale=0.85]{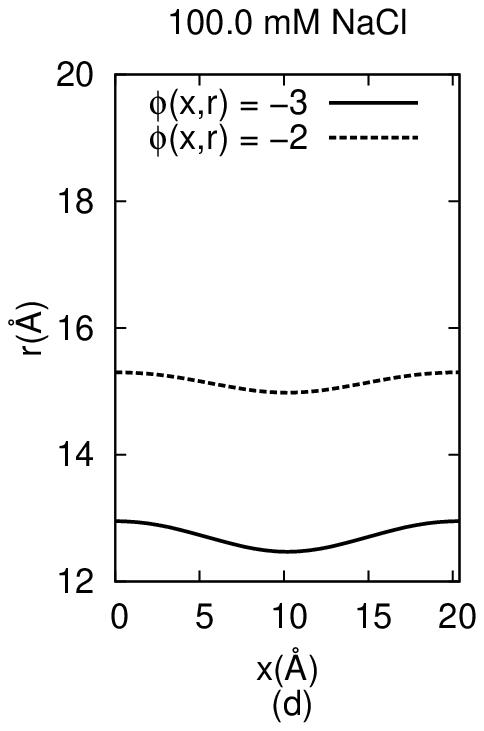}
\end{minipage}
\caption{\label{fig:potII}Potential contour lines at a distance $r$ perpendicular to a charged rod (model A) calculated by method II for different salt concentrations.}
\end{figure}
Figure \ref{fig:potII} depicts the potential contours around the charged rod in different salt concentrations for rod model A (Fig. \ref{fig:rod-models}) generated by method II calculations. The variables $r$ and $x$ are as for the previous subsection.\\ 
\begin{figure}[h!]
\includegraphics[scale=0.9]{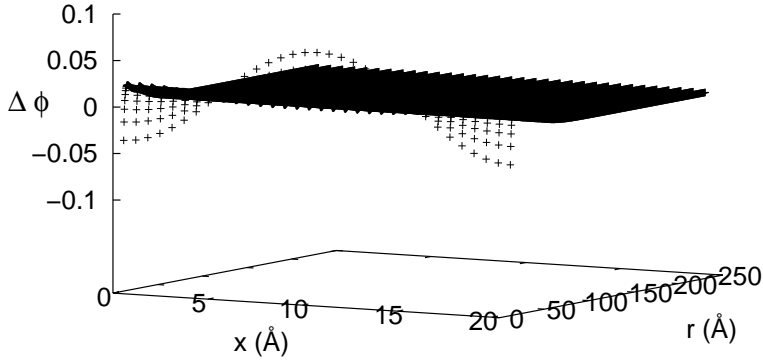}
\caption{\label{fig:Delta-I-II}Potential difference  $(\Delta \phi = \sub{\phi}{II} - \sub{\phi}{I})$ from the potential calculation for system of rod model A using method I and II. Salt concentration 100 mM.}
\end{figure}
Fig. \ref{fig:Delta-I-II} depicts the difference between the potential calculated by method I and method II for the model A rod system in NaCl salt concentration 100 mM. It will be observed that the potential difference are minute (of the order of 1 \% since $\phi \approx -3.6$ at low $r$). We attribute the source of error to (i) numerical differencing and (ii) interpolation.\\
\subsection{Radial Distribution Functions}
\begin{figure}[h!]
\hspace{1.5cm}
\includegraphics[scale=0.9]{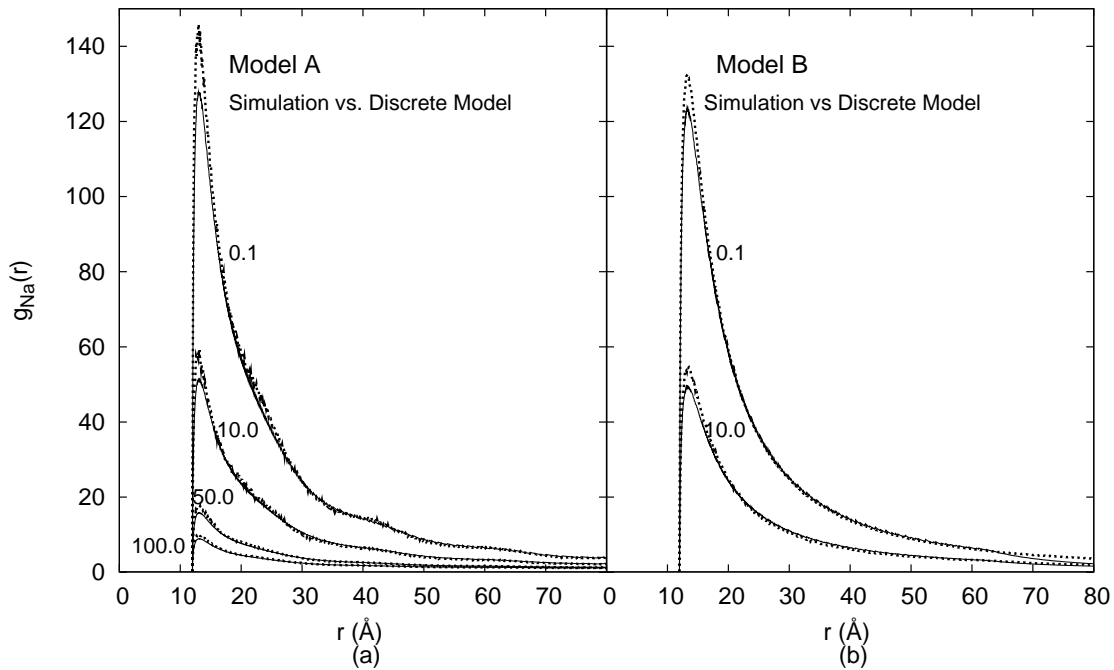}
\caption{\label{fig:rdf-AB}The theoretical rod charge$\--$Na RDF (solid lines) and simulation RDF (dotted lines) for rod model A (Fig. A) and rod model B (Fig. B). The theoretical potentials about the discretely charged rod are calculated by method I (Subsection \ref{sub:Method1}). The value 0.1; 10.0; 50.0 100.0 are the respective salt concentrations in millimolar.}
\end{figure}

From the theoretical potential data, the theoretical RDF is produced by the procedure of Section \ref{sec:get-rdf}. Fig. \ref{fig:rdf-AB} gives the RDF's of the Na$^+$ ions about the rod charge of rod model A (Fig. \ref{fig:rdf-AB}.a) and model B (Fig. \ref{fig:rdf-AB}.b) in different salt concentrations.\\
\noindent In Fig. \ref{fig:rdf-AB} we observe some oscillations of the RDF for the simulation and theoretical RDF for rod model A which are not observed for rod model B. We define oscillations have to be smooth and wavy slope of the graph. On the other hand the simulations also exhibits fluctuations in addition to the oscillations that occur in both. Little oscillations that occur at the simulation RDF of the rod model A has exactly the same period as the PBE RDF. The coincidence suggests that the principal cause of these oscillations in the simulation RDF is the nonuniform potential distribution parallel and perhaps perpendicular to the rod axis. For low salt concentration, there is a disagreement between the height of the first peak of the simulation and calculated RDF, both for rod model A and B. This disagreement reduces as the salt concentration increases. The height of the first RDF peak between the simulation for model A and B is also different. This difference, which also appears for the RDF of DNA polyelectrolyte simulation in Fig. \ref{fig:rdf-DNA}, indicates that the counterion distribution at the first layer about the charges of rod model A is denser than for the rod model B.\\
The above discussions indicates that discrepancies can occur when simulating a bunch of charges as equivalent to one single charge.\\
\section{Conclusion}
We have shown that the potential profile for model A and B is quite different, implying that criteria based on theoretical modelling should be used in conjunction with MD simulations. Previously this aspect has been ignored in detail and elaborate studies that might open to question the validity of the results and conclusions. Method I and II generate identical results for the potential profile although the algorithms are entirely different. We can therefore exploit these methods differently where one method might be more tractable than the other for a particular systems. The theoretical and simulation RDF is in very good agreement, except at the first peak at low salt concentrations. We believe that the equations we solved are incomplete, especially in relation to the screening effects that are not taken into account. Finally it cannot be overemphasized that theoretical criteria must be adopted in choosing appropriate charge and size of the particles involved in the molecular system that is simulated.\\
\section{Acknowledgement}
This work was funded by the University (UM) grant UMRG-RG077/09AFR and Government grant FRGSFP084/2010A. A.A.J.A is grateful for a research assistantship from UM.

\appendix
\section{Multigrid method for solving the cylindrical PBE}\label{appA}
\begin{figure}[h!]
\includegraphics[scale=0.9]{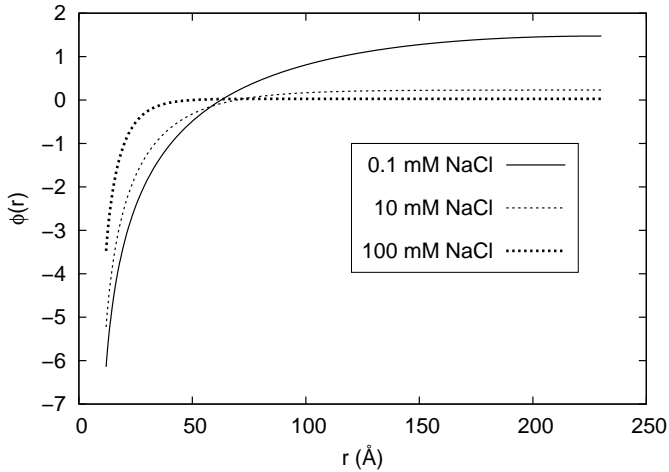}
\caption{\label{fig:MG-pot-conti}The PBE potential solutions around a continuously charged rod calculated by multigrid method. $r$ is the perpendicular distance to the rod axis. The rod model has continuous charge density $\delta=\sub{z}{e}/b= 1/1.7 \angstrom^{-1}$.}
\end{figure}

To generate the potential data from (\ref{eq:cylPB}), we apply the nested iteration multigrid method combined with the globally convergent Newton method (GCNM) for the smoothing process. In some cases, solving (\ref{eq:cylPB}) with relaxation or shooting method cannot reach convergence. We also found that the GCNM offers better stability than the nonlinear conjugate gradient method for the smoothing process. The code is largely derived from \citet{Pre92}, where we have modified the code, including replacing the original equation, the smoothing functions, and the boundary conditions (from Dirichlet to Neumann) in addition to expanding from 2-D calculation with equal number of grid for both axes to the 1-D and 2-D calculation where the number of grid at both axes can be different.\\ 
Following the notation in \citep{Pre92}, we define $\mathcal{L}(u)$ (see Eq. (19.6.22) at Section 19 in \citep{Pre92}) as the source term for (\ref{eq:cylPB}) such that
\begin{equation}
\mathcal{L}(u)= \frac{\partial^2u}{\partial x^2} + \frac{1}{x}\left (\frac{\partial u }{\partial x}\right) - \sum_{i=1}^s A_i\exp(-z_i u )
\end{equation}
where $u$ and $x$ denote  $\phi(r)$ and $r$ respectively.\\
The following is one of the modified portions of the file \emph{mgfas.c} from (\citet{Pre92}). This C file is the driver for the nonlinear multigrid code.\\
========================\\
\renewcommand{\familydefault}{\sfdefault}
for(jj=j; jj$>$= 2; jj$--$) \{			/* downward stoke of the V */\\
relaxGCNR(); /* GCNR (Smoothing function) */\\
//relaxNLCG(); /* Non-linear Conjugate Gradient (Smoothing function)*/\\
$\cdots$\\	
$\cdots$\\	
\}\\
dtmp1= iu[1][2]; \\
itmp1= renew\_continuity(); /* solve the coarsest grid solution */ \\
if(itmp1) printf("Coarsest grid solved, mid point \%f $-->$ \%f", dtmp1, iu[1][2]); \\
else iu[1][2] = dtmp1; /* no coarsest solution */ \\
for(jj=2; jj <= j; jj++) \{		/* Upward stroke of V */ \\
$\cdots$	 \\
$\cdots$	 \\
relaxGC\_NR();  \\
//relaxNLCG(); ~\\
\}\\ 
update\_neumannbc(); /* check and update at boundaries */\\
========================\\
\renewcommand{\familydefault}{\rmdefault}
Figure \ref{fig:MG-pot-conti} depicts the potential solutions of (\ref{eq:cylPB}) in different salt concentrations using the above method. We notice that the potential rapidly converges to zero at higher NaCl concentrations, which is to be expected. The von Neumann boundary conditions are also satisfied in the vicinity of the boundary.\\   
\\

\end{document}